\documentclass[twocolumn,floatfix,superscriptaddress,a4paper,showkeys,nofootinbib,reprint]{revtex4}
\usepackage[colorlinks=true,linktocpage=true,linkcolor=blue,citecolor=blue,allcolors=blue]{hyperref}
\usepackage{epsfig}
\usepackage{latexsym}
\usepackage[utf8]{inputenc}
\usepackage{xspace}
\usepackage{indentfirst}
\usepackage{enumitem}
\usepackage{color}
\usepackage{setspace}
\usepackage{lipsum}
\usepackage{graphicx}
\usepackage{todonotes}

\usepackage{amsmath}
\usepackage{amssymb}
\usepackage[english]{babel}
\usepackage{url}

\newcommand{\eq}[1]{\begin{align} #1 \end{align}}

\UseRawInputEncoding
\begin{document}
	
	\title{Critical point and Bose-Einstein condensation  in pion matter}
	\author{V. A. Kuznietsov }
	\affiliation{Physics Department, Taras Shevchenko National University of Kyiv, 03022 Kyiv, Ukraine}
	\author{O. S. Stashko}
	\affiliation{Physics Department, Taras Shevchenko National University of Kyiv, 03022 Kyiv, Ukraine}
	\author{O. V. Savchuk}
	\affiliation{Frankfurt Institute for Advanced Studies Frankfurt am Main, Germany}
	\author{M. I. Gorenstein}
	\affiliation{Frankfurt Institute for Advanced Studies Frankfurt am Main, Germany}
	\affiliation{
		Bogolyubov Institute for Theoretical Physics, 03680 Kyiv, Ukraine}

	\date{\today}
	\begin{abstract}
		\normalsize
		
		The Bose-Einstein condensation and the liquid-gas first-order phase transition are studied in the interacting pion matter. Two phenomenological models are used: The mean-field model and the hybrid model. Free model parameters are fixed by fitting the lattice QCD data on the pion Bose condensate density at zero temperature. In spite of some minor differences, the two models demonstrate an identical qualitative and very close quantitative behavior for the thermodynamic functions and electric charge fluctuations. A peculiar property of the considered models is an intersection of the Bose-Einstein condensation line and the line of the first-order phase transition at the critical endpoint.
	
	\end{abstract}
	
	\maketitle
	\section{Introduction}
	The Bose statistics \cite{Bose1924} and Bose-Einstein condensation (BEC) phenomenon \cite{Einst1925} were predicted almost hundred years ago. The BEC can take place in equilibrium systems of noninteracting bosons when a macroscopic part of all particles begins to occupy a single zero-momentum state. This was approximately realized experimentally in atomic systems at very small temperatures and particle number densities \cite{Anderson:1995gf, PhysRevLett.75.3969, PhysRevLett.75.1687, RevModPhys.71.463}.

	The BEC can also happen in condensed matter, nuclear physics, astrophysics, and cosmology (see, e.g., Refs. \cite{Satarov_2017, Begun:2006gj, Begun:2008hq, Strinati_2018, Nozieres:1985zz, PhysRevLett.101.082502, Chavanis:2011cz, Mishustin_2019, Padilla_2019}).In most of these situations, particle interactions should be taken into account. If both the attractive and repulsive interactions between particles are taken into account, the system reveals the first-order liquid-gas phase transition (FOPT) and the critical endpoint (CP). Therefore, the two phenomena — BEC and CP—can be expected for interacting bosons.
	
	Pions are three pseudo-scalar mesons, $\pi^+$, $\pi^0$, $\pi^-$,  that obey the Bose statistics, thus, an emergence of the BEC of pions is possible. The BEC can take place during the cooling of the early Universe \cite{Vovchenko_2021}, in the gravitationally bound pion stars \cite{Brandt_2018, Mannarelli_2019, andersen2018boseeinstein}, or as a non-equilibrium phenomenon in heavy-ion collisions \cite{Begun:2006gj, Begun:2008hq, Begun_2015}.  It has been also predicted to occur at large isospin (electric) chemical potentials \cite{Son:2000xc, Abuki:2009hx}. 
	This suggestion has been studied in 
	Quantum Chromodynamics 
	(QCD), a theory of strong interactions. 
	A presence of the BEC was  supported by recent first principle lattice QCD simulations \cite{Brandt:2017oyy, Brandt:2018bwq}. The Bose condensate (BC) of pions was observed at low temperature, $T<m$, and large electric (isospin) chemical potential $m <\mu<2m$, where $m$ is the pion mass (in what follows we use an approximate value $m=140$~MeV neglecting a small difference between the masses of neutral and charged pions).  In this specific region of the $(\mu,T)$ phase diagram the  QCD matter is expected in a form of the interacting pions. The heavier hadrons and/or quark-gluon degrees of freedom are expected to be suppressed at $T\ll m$ and not too large $\mu$.

	Various approaches to a description of the pion matter were developed: chiral perturbation theory~\cite{Adhikari:2019zaj,Adhikari:2020kdn}, linear sigma model~\cite{Andersen:2006ys,Andersen:2008qk}, Nambu-Jona-Lasinio 
	model \cite{PhysRevD.71.116001} or Polyakov-loop extended quark meson model \cite{PhysRevD.98.074016, Folkestad_2019}, functional renormalisation group~\cite{KAMIKADO20131044, SVANES201116}, hard thermal loops~\cite{PhysRevD.93.054045}, self-interacting mean field theory~\cite{Stashko_2021}, etc.

	Influence of particle interactions on the thermodynamic properties was also considered within the $S$-matrix formulation of statistical mechanics~\cite{Dashen:1969ep}. In particular, the attractive and repulsive interaction from  hadron-hadron scatterings were discussed in a number of works~\cite{Venugopalan:1992hy,Broniowski:2015oha,Savch2020,Fernandez-Ramirez:2018vzu,Dash:2018mep}. A possibility of the BEC in the pion system with zero chemical potential was also discussed using a Skyrme-like model including both attractive and repulsive interaction \cite{Mishustin_2019, Anchishkin_2019,Stashko_2021}. 
	
	A description of the repulsive and attractive interaction in statistical systems of hadrons is often performed in terms of the following phenomenological approaches: Mean-field approximation, effective-mass model, excluded volume approximation, etc. The mixtures of these approaches are also used. For example, a famous Walecka model \cite{Walecka:1974qa} for nuclear
    matter describes repulsive interactions in terms of the mean field $U(n)$ being a linear function of nucleon number density, and attractive interactions in terms of the effective nucleon mass. These different phenomenological models belong to the same universality class, the so-called “classical models”
	\cite{LL}
	 (the names “mean-field models” and “van der Waals-type models” are also used). These classical models lead to very similar description of the FOPT and CP (see, e.g., Ref. \cite{Poberezhnyuk:2017yhx}).  The models from the same universality class for the CP can however lead to rather different consequences for the BEC. This was discussed in Ref. \cite{Savch2020} where only the repulsive parts of interactions were taken into account.

	In Ref. \cite{PhysRevC.103.065201} the effective mass model with $\phi^4$ attractive and $\phi^6$ repulsive interaction was studied. It was found that both phenomena, FOPT and BEC,
	take place. An additional peculiar feature of the model was an observation that the CP belongs to the line of the BEC. How do these  features of the pion matter depend on the specific model used in Ref. \cite{PhysRevC.103.065201}? 

	This question motivates the present studies. We discuss two other phenomenological descriptions of the pion matter. The first model considers both attractive and repulsive interactions in terms of the mean field $U(n)$ depending on the pions number density. The second ``hybrid" model treats the repulsive interactions as the mean field, and the attractive interactions in terms of the effective pion mass $m^*(T,\mu)<m$.   We first fit the lattice QCD (lQCD) data at zero temperature and finite isospin chemical  potential $\mu$ to fix the model parameters. Then, thermodynamic functions and electric (isospin) charge fluctuations are calculated in the $(\mu, T)$ plane. 

	Intensive measures of the electric charge fluctuations, the scaled variance, skewness, and kurtosis, appear to be very sensitive to a presence of the CP and the BEC.
	The paper is  organized as follows.  The ideal Bose gas of pions is considered 
	in Sec. \ref{IBG}. The two phenomenological models of the interacting pion matter are discussed and compared in Sec. \ref{IntBG}. 
	Summary in Sec. \ref{SUM} closes the paper.
	
	\begin{figure*}
		\centering
		\includegraphics[width=.95\textwidth]{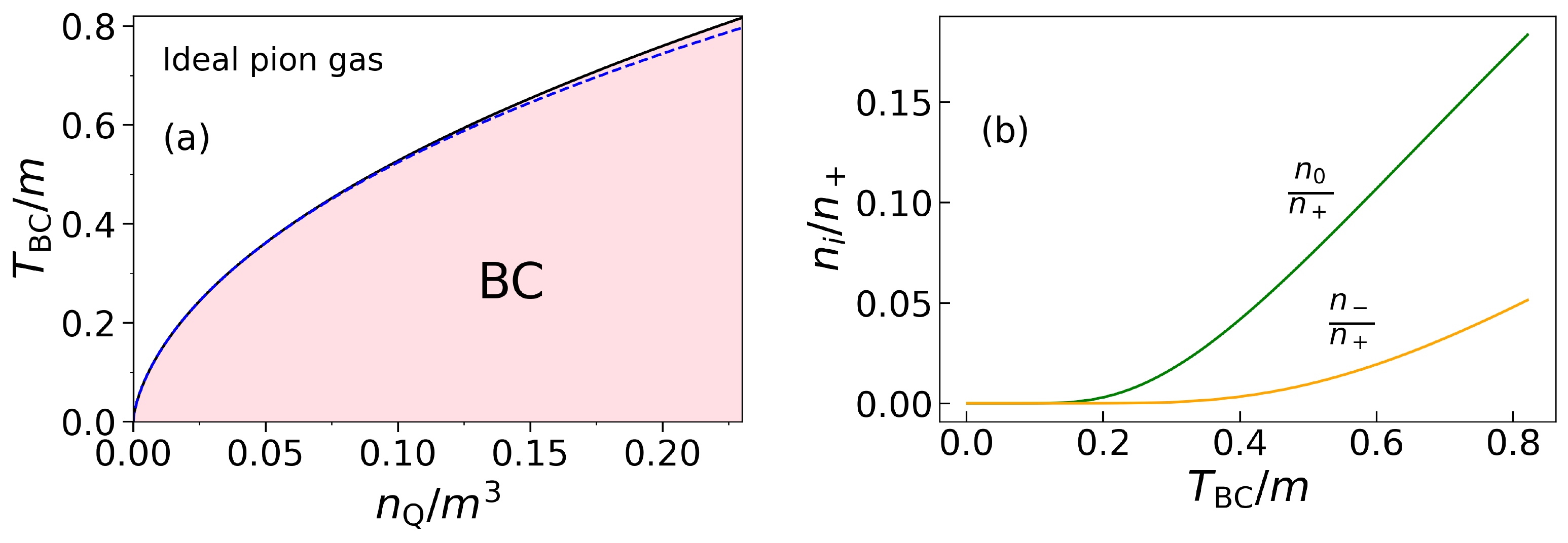}
		
		\caption{\label{fig:1}(a): A line of the BEC $T_{\rm BC}$ for the ideal pion gas as a function of electric charge density $n_Q$ is shown by a solid line. A dashed line shows  the same function when the $\pi^-$ presence is neglected. (b): Ratios of the $\pi^0$ and $\pi^-$ number densities to that of $\pi^+$ in the IdPG on the BEC line as functions of $T_{\rm BC}$. 
		}
	\end{figure*}

	\section{Ideal pion gas}\label{IBG}
	
	The ideal pion  gas (IdPG) is described  in the grand canonical ensemble by the  pressure function \cite{GNS}
	\eq{
	&	p_{\rm id}(T,\mu)  =\sum_{i=+,0,-}	p_{\rm id}(T,\mu_i) %
		\nonumber \\
		& = 
		-\sum_{i=+,0,-}\frac{1}{2\pi^2}\int\limits_{0}^{\infty} dk\,k^2 \log\left[1-\exp\left(\frac{\mu_i-\sqrt{k^2+m^2}}{T}\right)\right]
		\nonumber \\
		& =
		\sum_{i=+,0,-}\frac{1}{6\pi^2}\int\limits_{0}^{\infty} dk  \frac{k^4}{\sqrt{k^2 + m^2}}f_k(T,\mu_i)~,
	}
	where integration by parts was used.

	The chemical potential $\mu$ corresponds to the electric charge conservation, 
	\eq{
		\mu_+ = \mu, \quad \mu_0 = 0, \quad \mu_- = -\mu~,
		\label{mu-i}
	}
	and
	\begin{equation}
	f_k(T,\mu_i) = \left[\exp{\left(\frac{\sqrt{k^2 + m^2} - \mu_i}{T}\right)} - 1\right]^{-1},
	\end{equation}
	is the Bose momentum distribution.
	The particle number densities of the $i$-th sort of pions equals to
	\eq{		n_i^{\rm id}(T,\mu_i) = \left(\frac{\partial p_{\rm id}}{\partial \mu_i}\right)_T= \frac{1}{2\pi^2} \int\limits_{0}^{\infty} dk \ k^2 f_k(T, \mu_i)~. \label{ni}
	}
	The electric charge density is calculated as	
	\eq{
		n_Q(T,\mu)=\left(\frac{\partial p}{\partial \mu}\right)_T = n_+^{\rm id}(T,\mu) - n_-^{\rm id}(T,-\mu)~,
	}
	and the number density of all pions as
	\eq{
		& \sum_{i=+,0,-} n_i^{\rm id}(T,\mu_i) 
		\nonumber\\
		&=\frac{Tm^2}{2\pi^2}\sum_{l=1}^{\infty}\frac{1}{l}K_2(lm/T)
		\left[1+2\cosh\left(\frac{l\mu}{T}\right)\right]~, \label{n}
	}
	where $K_2$ is the modified Bessel function.
	
	The inequality $|\mu | \le m$ should be satisfied in the IdPG. An onset of the BEC occurs at $|\mu | \rightarrow  m-0$. The condition $n_Q=n_Q(T,|\mu |=m)$ defines then a line $T_{\rm BC}=T_{\rm BC}(n_Q)$  
	denoted further  as the BEC line.    
	Under this line at $ T<T_{\rm BC}(n_Q)$ there is a region with the nonzero BC.
	In what follows we consider $\mu \ge 0$. It leads to $n_+\ge n_0\ge n_-$ and $n_Q=n_+-n_-\ge 0$, where $n_+,n_0$, and $n_-$ denote the number densities of $\pi^+,~\pi^0$, and $\pi^-$, respectively. The results for $\mu \le 0$ can be obtained by interchanging $\pi^+$ and $\pi^-$.
	A line of the BEC $T_{\rm BC}(n_Q)$ for the IdPG corresponds to $\mu\rightarrow m-0$. It is shown in the $(n_Q,T)$ plane  in Fig.~\ref{fig:1} (a)\footnote{In figures of the present paper we use dimensionless variables with the pion mass $m$ as the  energy scale.}. 
	At this line  there is an onset of the $\pi^+$ BEC. Under this line  nonzero values of the $\pi^+$ BC are formed. The total pion number density (\ref{n}) should be then modified at $T<T_{\rm BC}(n_Q)$,
	\eq{
		n = \sum_{i=+,0,-} n_i^{\rm id}(T,\mu_i) + n_{\rm BC}^+~,
	}
	where 
	$n_{\rm BC}^+\ge 0$ corresponds to the BC of $\pi^+$. 
	At $T\rightarrow 0$ all thermal densities (\ref{ni}) vanish. Thus, $n= n_Q= n_{\rm BC}^+$ at $T=0$, i.e., at zero temperature the pion system consists from the pure BC of $\pi^+$.
	A dashed line in Fig.~\ref{fig:1} (a)  shows the BEC line for only one sort of pions, i.e., for $\pi^+$ It gives a  good approximation of the general case with all three sorts of pions. This is because 
	 of  $n_-\ll n_+$ on the BEC line $T_{\rm BC}(n_Q)$.
	Therefore, an analytic behavior of the $T_{\rm BC}(n_Q)$ at $T_{\rm BC}/m \ll 1$corresponds approximately to the well-known textbook result for one sort of nonrelativistic bosons \cite{LL}:
	\eq{ 
		T_{\rm BC}(n_Q)\cong \frac{2\pi}{m}\,\left(\frac{ n_Q}{\zeta(3/2)}\right) ^{2/3}~, \label{Tbc}
	}
	where $\zeta(3/2)\cong 2.612$ is the Riemann zeta function. 
	Corrections to Eq.~(\ref{Tbc}) from $\pi^-$ are small as seen from Fig.~\ref{fig:1} (a). Relativistic corrections to Eq.~(\ref{Tbc}) at $T/m\ge 1$ are considered in Ref.~\cite{Begun:2008hq}. 
	The ratios $n_0/n_+$ and $n_-/n_+$ along the BEC line are shown in Fig.~\ref{fig:1} (b).  The inequalities $n_+\gg n_0\gg n_-$
	remain valid on the BEC also for the interacting pion matter.
	An intensive measure for fluctuations of the electric charge $Q=N_+ -N_-$ is the scaled variance 
	\eq{\omega_Q =\frac{\langle Q^2\rangle~-~\langle Q\rangle^2}{\langle Q\rangle }~, \label{Q}
	}
	where $N_+$ and $N_-$ are the total numbers of $\pi^+$ and $\pi^-$, respectively, and  $\langle \ldots \rangle$ denotes the grand canonical averaging.
	In the IdPG  it takes the following form
	\eq{	&\omega_{Q}^{\rm id} = \frac{T}{n_Q}\left(\frac{\partial n_Q}{\partial \mu}\right)_T  = \frac{n_{\rm id}(T,\mu) + 
			n_{\rm id }(T,-\mu)}{n_Q} \nonumber \\
		& + \frac{1}{2\pi^2n_Q} \int_{0}^{\infty} dk\,k^2  ~
		\left[f^2_k(T,\mu)  + f^2_k(T,-\mu) \right]~. \label{omegaQ}
 	}
	When $\mu=m$ the function $f_k(T,\mu=m)$ is proportional to $k^{-2}$ at  $k/m\ll1$. The integral $\int_0^\infty k^2 dk f_k^2(T,\mu=m)$ becomes thus divergent at the lower limit as $k^{-2}$. Therefore, the scaled variance $\omega_Q$ becomes infinite on the BEC line.
	
	At $\mu\rightarrow m-0$, the  electric charge fluctuations show an anomalous behavior 
$\omega_Q \propto V^{1/2}$ in the finite volume $V$. This leads to
infinite values of $\omega_Q$ on the BEC line \cite{Begun:2006gj,Begun:2008hq}:
$\omega_Q\propto (m- \mu)^{-1/2}$ at $\mu\rightarrow  m-0$
in the thermodynamic limit $V\rightarrow \infty$.  It	happens due to the BEC of $\pi^+$, and is rather similar to a behavior of the ideal Bose gas for one sort of particles.
 Under the BEC line an additional contribution $\omega_C$ from  the BC  $n_{\rm BC}^+> 0$ should be added to $\omega_Q$ (\ref{omegaQ}).
 It behaves as \cite{Begun:2008hq}
	\eq{\omega_C=V\frac{\left(n_{\rm BC}^+\right)^2}{3n_Q}}
in the large finite system and goes to infinity in the thermodynamic limit $V\rightarrow \infty$. 

	It should be noted that the scaled variance (\ref{Q}) can not be applied at $\mu\rightarrow 0$. It becomes meaningless in this limit as $\langle Q\rangle \rightarrow 0$. At $\langle N_+-N_-\rangle =0$ the other fluctuation measure $\langle (N_+-N_-)^2\rangle /
	\langle N_++N_-\rangle$ is usually used to describe the charge fluctuations.

	\begin{figure*}
		\includegraphics[scale=.5]{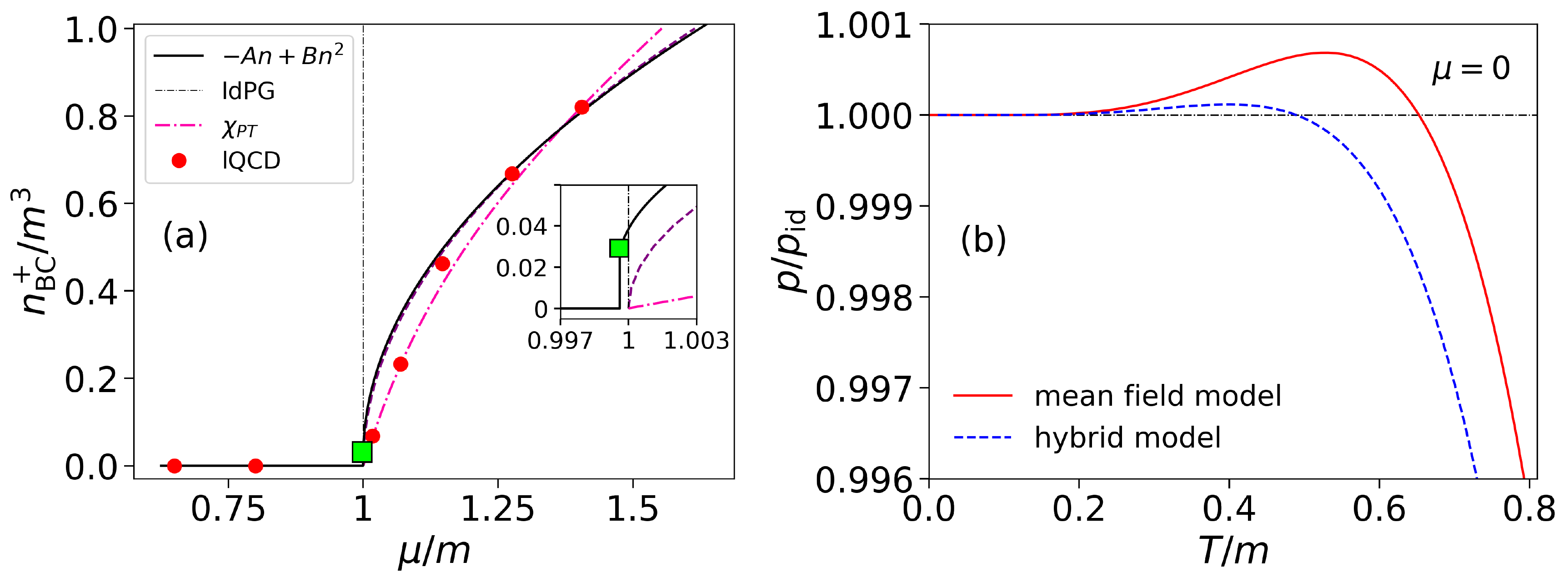}
		\caption{\label{fig:2}(a): A fit of the LQCD data   \cite{Brandt:2017oyy,Brandt_2018} of the 
		$n_{\rm BC }^+$  at $T=0$ as a function of $\mu$ in the mean-field model. A solid line shows the results with $A$ and $B$ given by Eq.~(\ref{AB}), dashed line corresponds to $A=0$ when the attractive interactions are neglected, and a dotted vertical line presents the IdPG results, i.e., when both $A=0$ and $B=0$. The leading-order of the chiral perturbation theory $\chi_{\rm PT}$ from Ref.~ 
		\cite{Son:2000xc}
		is  shown by a dashed-dotted line. A green box corresponds to the $\mu_0$ and $n_0$ values of the pion matter ground state.  (b): The ratio of $p$ to $p_{\rm id}$ as functions of $T$ for $\mu=0$. 
			A solid line corresponds to the mean-field model for $p$ with the parameters (\ref{AB}), a dashed line to the hybrid model discussed below.
			}
	\end{figure*} 
	
	\section{Interacting pion Gas}\label{IntBG}
	In this section, two phenomenological models for the repulsive and attractive interactions in the pion matter are considered. 
	\subsection{Mean-field model}

	The mean field model for the system of interacting pions is given by the following set of self-consistent equations:
	\eq{
		p(T,\mu) &=\sum_{i=+,0,-} p_{\text{id}}(T,\mu_i^*) + \int\limits_{0}^{n}dn^{\prime}n^{\prime}\frac{dU(n^{\prime})}{dn^{\prime}}~,\label{p1} \\
		n(T,\mu) &= \sum_{i=+,0,-} n_i^{\text{id}}(T, \mu_i^*) + n_{\text{BC}}^+~, \label{n-1} \\
		\mu^*_i & = \mu_i - U(n)~, \label{mui*} 
	}
	where $U(n)$ is the mean field that describes pion interactions, $\mu_i$ in Eq.~(\ref{mui*}) is given by Eq.~(\ref{mu-i}), and    $n_{\rm BC}^+ $  in Eq.~(\ref{n-1}) is the BC density of $\pi^+$ that can attains non-zero values when the BEC condition, $\mu^*=m$, is fulfilled.  A second term in the right hand side of Eq.~(\ref{p1}) corresponds  to the so-called 
	excess  pressure that makes Eqs.~(\ref{p1})-(\ref{mui*})  
	to be thermodynamically self-consistent (see, e.g.,  Ref. \cite{Anchishkin_2015}).
	The mean field potential $U(n)$ will be  taken in the following form:
	\begin{equation}\label{Un}
	U(n) = -An + Bn^{2}, \quad A > 0, \ B > 0~,
	\end{equation}
	where constants $A$ and $B$ correspond to the attractive and repulsive interactions, respectively.  
	In the model with mean field the BEC can take place at both $\mu <m$, for $U(n) <0$, and $\mu >m$, for $U(n)>0$. For the potential $U(n)$ given by Eq.~(\ref{Un}) these conditions correspond to small and large values of $n$, respectively.

	At $T=0$ the pion system can only exist in the form of the $\pi^+$ BC. A condition of the BEC, $\mu-U(n_{\rm BC}^+)=m$, leads to the following
	solution for $n_{\rm BC}^+$:  
	\begin{equation}\label{nBC0}
	n_{\rm BC}^+(T=0,\mu) = \frac{A + \sqrt{A^2 + 4B(\mu - m) }}{2B}~.
	\end{equation}
	The system pressure at $T=0$ is given by a second term in the right hand side of Eq.~(\ref{p1}):
	\eq{\label{pT0}
		p=-\frac{A}{2}\,\left(n_{\rm BC}^+\right)^2+\frac{2B}{3}\,\left( n_{\rm BC}^+\right)^3~.
	}
	At small $\mu$ the system exists in the ``gaseous'' phase with $n=0$ and $p=0$. At some $\mu=\mu_0$ there is the FOPT. According to the Gibbs criteria  the pion BC density jumps to the ``liquid'' phase with $n_{\rm BC}^+ = n_0 >0$  and the pressure $p= 0$. 
	This FOPT takes place  at
	\eq{
		n_{\rm BC}^+=\frac{3A}{4B}~\equiv n_0, \label{n0}
	}
	where the liquid pressure (\ref{pT0}) equals to zero, and the chemical potential $\mu_0$ found from Eq.~(\ref{nBC0}) is
	\eq{\label{mu0}
		\mu_0=m -\frac{3A^2}{16B}~,
	}
	which defines also the ground state binding energy per pion 
	\eq{\label{W}
		W \equiv \frac{\varepsilon}{n_0} -m~=~-\frac{3A^2}{16B}~.
	}
	
	At $\mu>\mu_0$ the BC $n_{\rm BC}^+$ as a function of $\mu$ is defined by Eq.~(\ref{nBC0}).
	In Fig.~\ref{fig:2} (a) the fit of the Monte Carlo lattice data for $n_{\rm BC}^+$ at $T=0$ with Eq.~(\ref{nBC0}) is presented. The fitting parameters $A$ and $B$ are fixed as
	\eq{\label{AB}
		A = 0.05~m^{-2}~,~~~ B=1.30~m^{-5}~. 
	}
	With these parameters one finds:
	\eq{\label{GS1}
		n_0 & \cong 0.029~m^3\cong 0.01~{\rm fm}^{-3}~,\\~~~
		W & \cong - 0.00036 ~m\cong -0.05~{\rm MeV}~. 
	}

	This  ground state of the pion BC looks  rather rarefied and  weakly bounded when it is compared with the ground state of the nuclear matter, $n_0^{\rm nuc}\cong 0.16$\,fm$^{-3}$ and $W_{\rm nuc}\cong -16\,$MeV.

	\begin{figure*}
		\includegraphics[width=0.95\textwidth]{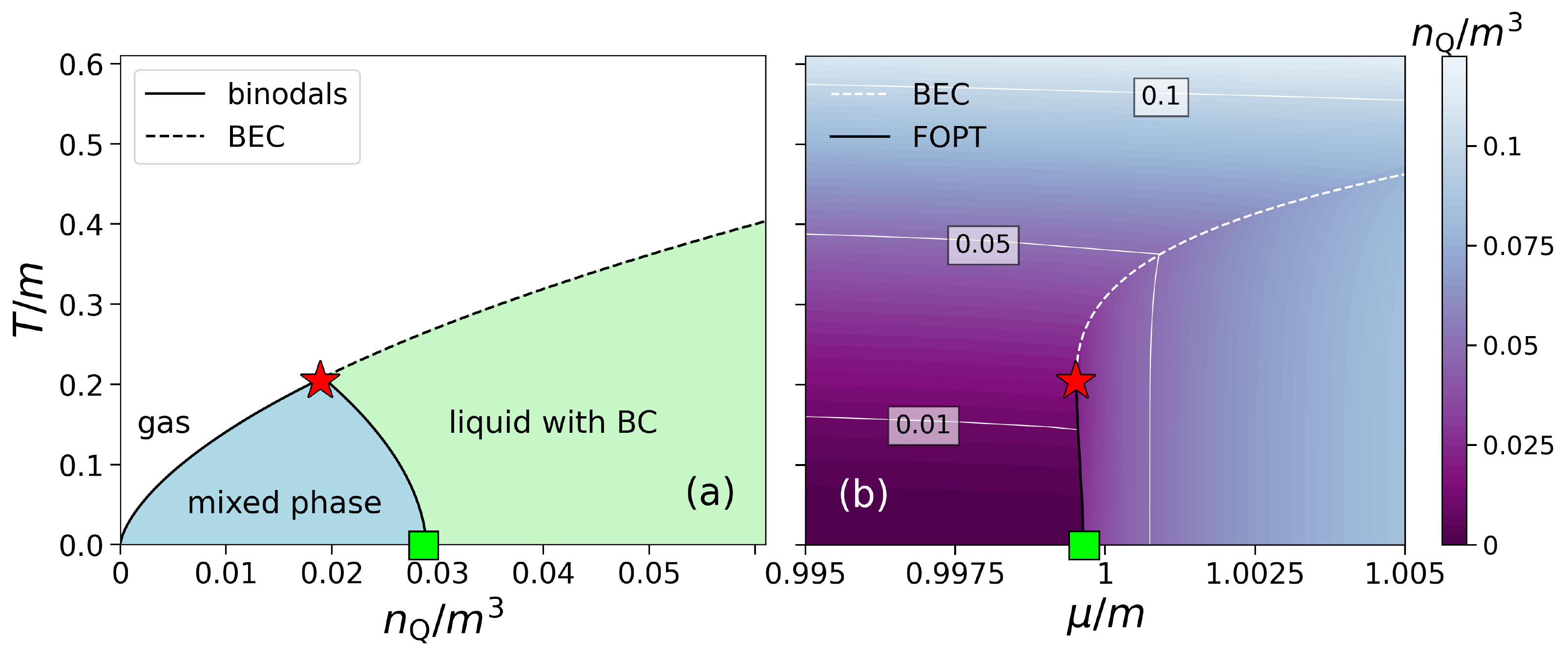}		
		\caption{\label{fig:3}(a) The phase diagram of the pion matter in the mean-field model with parameters (\ref{AB}) on  the $(n_Q,T)$ plane. The left and right binodals  are shown by solid lines, the BEC line by dashed line.  A green box denotes a ground state with $n_Q=n_0$ and  $T=0$, and a red star denotes the CP with $T_c\cong 0.21\,m$. (b): The same as in (a), but for the $(\mu,T)$ plane.}		
	\end{figure*}  
	
	\begin{figure*}
		\includegraphics[scale=.5]{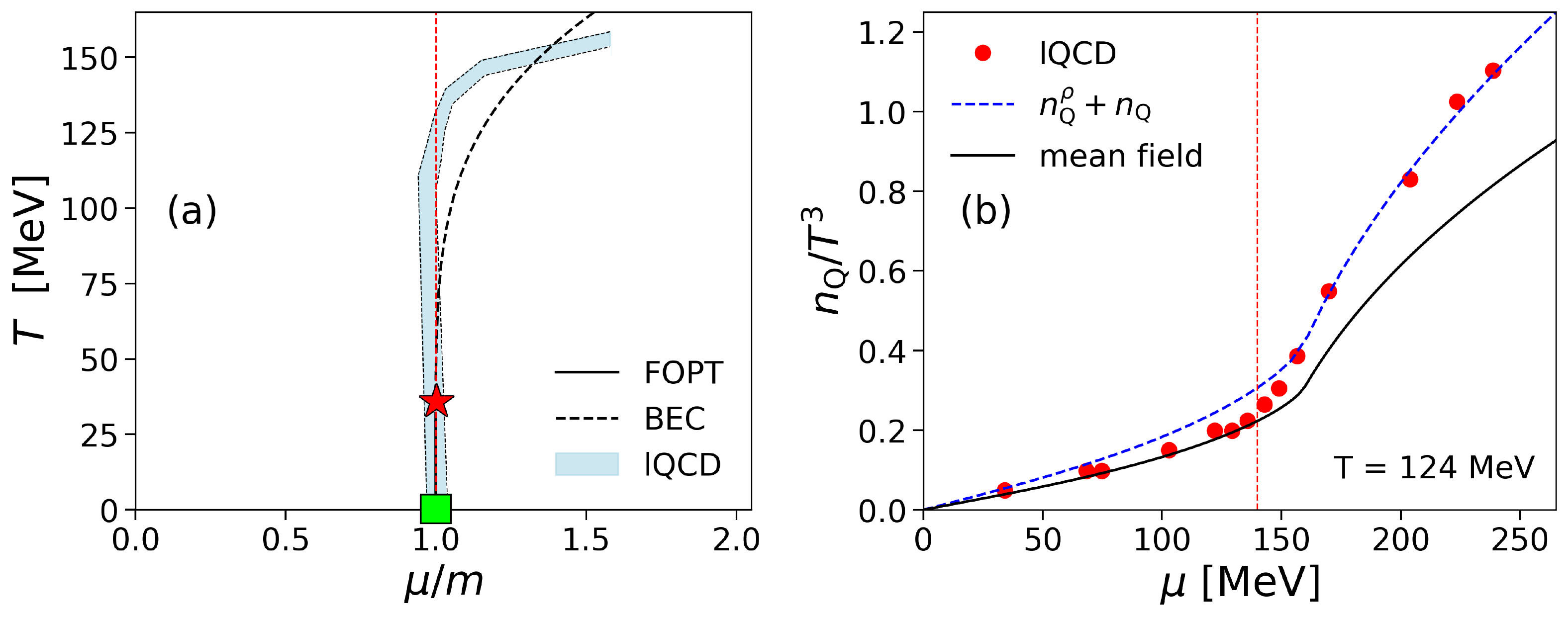}
		\caption{\label{fig:4}(a): A comparison of the BEC line in mean field model at $T > 0$ with the lQCD data \cite{BBrant_2018}. (b): The charge density $n_Q$ from the mean-field model and the  lQCD data  \cite{2019} at $T = 124$ MeV. A dashed line includes additional contribution (\ref{rho}) to $n_Q$ from $\rho^\pm$ mesons. Vertical dotted lines indicate $\mu = m$.}
	\end{figure*} 
	
	In Fig.~\ref{fig:2} (b) the ratio of the pressure of interacting pion gas to that of the IdPG  is shown by a solid line as a function of $T$ at $\mu=0$. The effects of the repulsive interaction suppress the pion pressure, $p<p_{\rm id}$, 
	and they are seen at large $T$. The repulsive effects are still rather moderate and correspond approximately to the excluded volume corrections with rather small  hard-core pion radius  $r\cong 0.13$~fm. Tiny effects of the attractive interactions are only seen at small $T$ where $p>p_{\rm id}$. Note that our modelling concerns the non-resonance part of the pion-pion interactions. These interactions contribute to the $n_{\rm BC}^+(\mu)$  at $T=0$ presented in Fig.~\ref{fig:2} (a). At $\mu=0$ a contribution of these non-resonance residual  pion interactions to the thermodynamic  functions are small.
	There are almost no chances to observe them in hadron statistical {\it equilibrium}  models  using the data on heavy-ion collisions. 
	This is because of $|\mu| \ll m$ in the equilibrium systems created in heavy ion collisions.
	Previous suggestions to observe the BEC in heavy ion collisions assumed a large values of $\mu\approx m$ for all three types of pions due to chemical {\it non-equilibrium} effects \cite{Begun:2006gj,Begun:2008hq} .

	The  mean-field model with free parameters $A$ and $B$ was discussed in Ref.~\cite{Stashko_2021} for $\mu=0$.  Several interesting phenomena including the BEC can take place in this case  due to large attractive interactions, $A\ge A_{\rm cr}=2(Bm)^{1/2}$. The required large values of the parameter $A$  are however fully unrealistic ones, $A_{\rm cr}$ is about 45 times larger than the $A$ value given by Eq.~(\ref{AB}).  These large values $A\ge A_{\rm cr}$  are  in a strong  contradiction with lattice data at $T=0$.

	In contrast to  $\mu=0$ case, at large	$\mu\cong m$ and $\mu>m$ the {\it small} pion interactions are however crucially important. A vertical dotted line $\mu=m$ in Fig.~\ref{fig:2} (a) presents a behavior $\mu=m$ in the IdPG. As seen from Fig.~\ref{fig:2} (a) the IdPG behavior is far away from the lattice data. A dashed line in Fig.~\ref{fig:2} (a) presents the model results at $A=0$, i.e., when only repulsive interactions are included. The both fits, with $A>0$ and $A=0$, are of the  similar quality with $\chi^2$ per degree of freedom (dof) $ \cong  2$. This fact means a dominant role of the repulsive interactions. The lattice data at $T=0$ presented in Fig.~\ref{fig:2} (a) can not give indisputable estimates of the parameter $A$. Similar conclusions were made in Ref.~\cite{PhysRevC.103.065201}. 

	The lattice data shown in Fig.~\ref{fig:2} (a) defines the value of parameter $B$ rather accurately, but give only some restrictions from above for $A$. 
	 Thus, a presence or absence of the FOPT remains as an open question.	   
	In Fig.~\ref{fig:2} (a) a dashed-dotted line presents the $\pi^+_{\rm BC}$ calculated in the leading-order chiral perturbation theory $\chi_{\rm PT}$ in Ref.~\cite{Son:2000xc}. This model being in a good agreement with the lattice data for $\pi^+_{\rm BC}$ at $T=0$ does not include the FOPT and predicts a second order phase transition.

	In the mean-field model the BEC line is defined by the condition 
	$\mu = m+U(n)$.  At  $U(n)>0$, one thus observes   $\mu > m$.
	The BEC line $T_{\rm BC}(n) $ is approximately a universal function of particle number density\footnote{This is an exact result for one sort of bosons in the mean-field model (see, e.g. Ref.~\cite{Satarov_2017}).}
	and it remains the same as for the IdPG where $U(n)=0$. This  is valid 
	for 
	$U(n)>0$ in the  region with no  FOPT. 
	At $A=0$ there is no FOPT and the BEC line $T_{\rm BC}(n_Q)$ look almost the same as for the IdPG shown in Fig.~\ref{fig:1} (a).  
	A behavior of the electric charge fluctuations $\omega_Q$ (\ref{omegaQ}) is however drastically changed on the BEC line and under this line due to a presence of the repulsive mean-field interactions. As will be seen below, in contrast to the IdPG, the scaled variance $\omega_Q$  
    becomes finite at all $n_Q>0$ due to the repulsive interactions.

	The phase diagram of the interacting pion gas is shown in Fig.~\ref{fig:3}. 
	For $A>0$ and $B>0$ the considered mean-field model reveals the FOPT.  
	The mixed phase region is constructed by the standard Gibbs procedure: when several solutions exist for the pressure function $p$ at the same $T$ and $\mu$ values, the physical solution corresponds to that with larger value of $p$.
	The line of the FOPT in the $(\mu,T)$ plane corresponds to the equal pressures of {\it gaseous} small density and {\it liquid} large density solutions.
	
	Figures \ref{fig:3} (a) and (b) present the  phase diagram on  the $(n_Q,T)$ and $(\mu,T)$ planes,  respectively. 
	A line of the FOPT in the $(\mu,T)$ plane starts at $\mu=\mu_0$ and $T=0$, and it ends at the CP  $T=T_c$ and $\mu=\mu_c$. The ground state point and CP point for the model parameters (\ref{AB}) are shown in Fig.~\ref{fig:3}  by a green box and red star, respectively.   
	A region of the mixed gas-liquid phase is bounded by the left and right binodal curves that are  shown in Fig.~\ref{fig:3} (a).
At  $T<T_c$, the pion system inside the mixed phase is an inhomogeneous mixture  of a rarefied gas and a dense liquid. The electric charge density $n_Q$ of the pion matter  
is then given by a linear combination of the gaseous phase with $n_Q^{\rm g}<n_Q$ lying on the left binodal, and the liquid phase with  $n_Q^{\rm l}>n_Q$ lying on the right binodal. The right  liquid binodal of the mixed state includes $n_{\rm BC}^+> 0$ while the  conditions for the BEC are not fulfilled in the left gaseous binodal. At $T\rightarrow 0$, one obtains $n_Q^{\rm g}\rightarrow 0$ and
$n_Q^{\rm l}=n_{\rm BC}^+$, i.e., the gaseous phase is vanished, and the liquid phase consists of the $\pi^+$ BC.

	A peculiar property of the model is a position of the CP
	lying on the BEC line. This is similar to the results of Ref.~\cite{PhysRevC.103.065201} and  is a consequence of smallness of the attractive forces in the pion matter. We checked that for $A$ larger than some critical value the BEC line intersects a line of the  FOPT at the triple point  $T=T_{\rm tr}<T_c$. 	Such a behavior was found for the interacting $\alpha$-matter in Ref.~\cite{Satarov_2017}.
	
	 We compare  the mean-field model with the lattice QCD results 
	\cite{BBrant_2018,2019} 
	at $T>0$. Figures \ref{fig:4} (a) and (b)
	present, respectively, a position of the BEC line in the $(\mu,T)$-plane and
	$n_Q$ as a function of $\mu$ at $T=124~$~MeV. One observes a good agreement with the lattice QCD data  in Fig.~\ref{fig:4} (a)  at not too large $T$.
	At $T>100$~MeV meson resonances give a substantial contribution to the thermodynamic observables. These degrees of freedom are  absent in the present version of the mean field-model. This point will be a subject for the future studies. To estimate 
	the resonance contribution to $n_Q$ we calculate the $n_Q^\rho$ value that comes from a presence of non-interacting $\rho^\pm$ mesons in the pion system:
	\eq{\label{rho}
&	n_Q^\rho = n_\rho^+~-~ n_\rho^- \nonumber \\ 
& =~ \frac{g_\rho}{2\pi^2}\int_0^\infty k^2dk
	\left[f_k(T,\mu; m_\rho)- f_k(T,-\mu; m_\rho)\right] ,
	}
	where $g_\rho=3$ and $m_\rho\cong 775$~MeV.
	A dashed line in Fig.~\ref{fig:4} (b) presents a sum
	of $n_Q$ value in the interacting pion system and $n_Q^\rho$ contribution (\ref{rho}) from the ideal $\rho$-meson gas.

	We introduce the quantities  ($i=+,0,-$)
	\eq{\label{wi}
		\omega_i  =  1 + \frac{1}{2\pi^2n_Q} \int_{0}^{\infty} dk\,k^2 f^2_k(T,\mu_i^*)~, %
	}
	that  are equal to the scaled variances of the IdPG
	$\omega_i^{\rm id}$ , but with shifted chemical potentials $\mu_i\rightarrow \mu_i^*$.  
	\begin{figure*}
		\includegraphics[width=.92\textwidth]{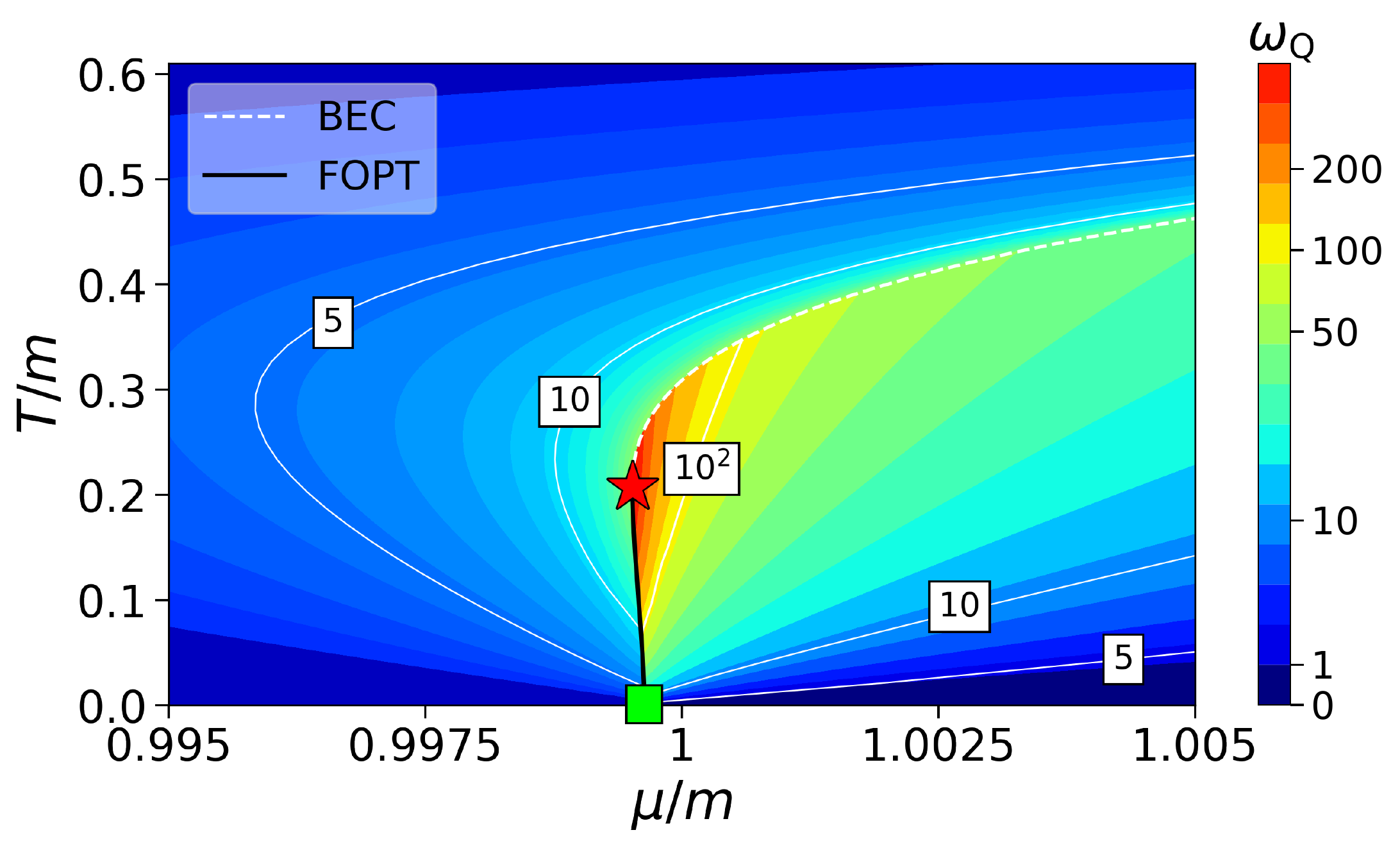}		
		\caption{\label{fig:5} The scaled variance $\omega_Q$  (\ref{omegaQ2}) on the $(\mu, T)$  plane in mean-field model.
			The lines,  green box and red star symbols are the same as in Fig.~\ref{fig:3}.
		}
\end{figure*} 
	One finds an explicit expression for $\omega_Q$ (see the Appendix):
	\eq{\label{omegaQ2}
		&\omega_Q =\frac{T}{n_Q}\,\left(\frac{\partial n_Q}{\partial \mu}\right)_T = 
		\frac{n_+\omega_+ + n_-\omega_-}{n_Q} \nonumber  \\ %
		&~- \frac{1}{n_Q}\,\frac{(dU/dn)\,(n_+\omega_+-n_-\omega_-)^2}{T+(dU/dn)\,\left(n_+\omega_+ + n_-\omega_-+n_0\omega_0\right)},
	}
	where $n_i\equiv n_i^{\rm id}(T,\mu_i^*)$.
	
	Approaching the BEC line, i.e., at $\mu^*_+\rightarrow m-0$,	the $\omega_i$ values demonstrate the following behavior:  $\omega_+  \rightarrow \infty$ while both $\omega_0$ and $\omega_-$ remain finite. This leads to the result on the BEC line:
	\eq{
		\label{omegaQ3}
		\omega_{Q} = \frac{T}{n_Q} \left[\frac{d U}{d n}\right]^{-1} + \frac{n_0 \omega_0 + 4n_- \omega_-}{n_{ Q}}.
	}

	If $A=0$ the value of $dU/dn$ is always positive at $n>0$. Thus, in contrast to the IdPG,
	the scaled variance $\omega_Q$ on the BEC line becomes finite due to the repulsive interactions. This conclusion remains also valid for $A>0$  at $T_{\rm BC}(n_Q)> T_c$ where  $dU/dn>0$.  It can be also shown that the scaled variance $\omega_Q$ is continuous across the BEC line.

	For $A>0$ the CP becomes the end point of both the FOPT line and the BEC line.
	When $T\rightarrow T_c$ along the BEC line   the  first  term in the right hand side of Eq. (\ref{omegaQ3}) goes to infinity, and, thus, the second finite term
	gives a negligible contribution to $\omega_Q$.  
	A relative contribution of this second term increases monotonously 
	with increasing $T_{\rm BC}(n_Q)$ along the BEC line. It still remains small for the discussed region of the system temperature. 
	
	Note that  $\omega_0\cong 1$, $\omega_-\cong 1$, and $n_0 < n_+$, $n_-\ll n_+$. 
	At  $T_c<T_{\rm BC}(n) < 0.8\,m$ the relative contributions to $\omega_Q$ (\ref{omegaQ3}) from each terms, $n_0\omega_0/n_Q$ and $4n_-\omega_-/n_Q$,
	are smaller than 3\%.
	Neglecting these small 
	contributions to $\omega_Q$ one obtains  
	an expression 

	\eq{\label{omegaQ4}
		\omega_Q
		\cong  \frac{T}{n_+}\,\left[ \frac{dU}{dn_+} \right]^{-1}.
	}
	It corresponds to the mean field model result for one particle species $n=n_+$
	(see, e.g., Ref.~\cite{Savch2020}).

    \begin{figure*}
		\includegraphics[width=0.90\textwidth]{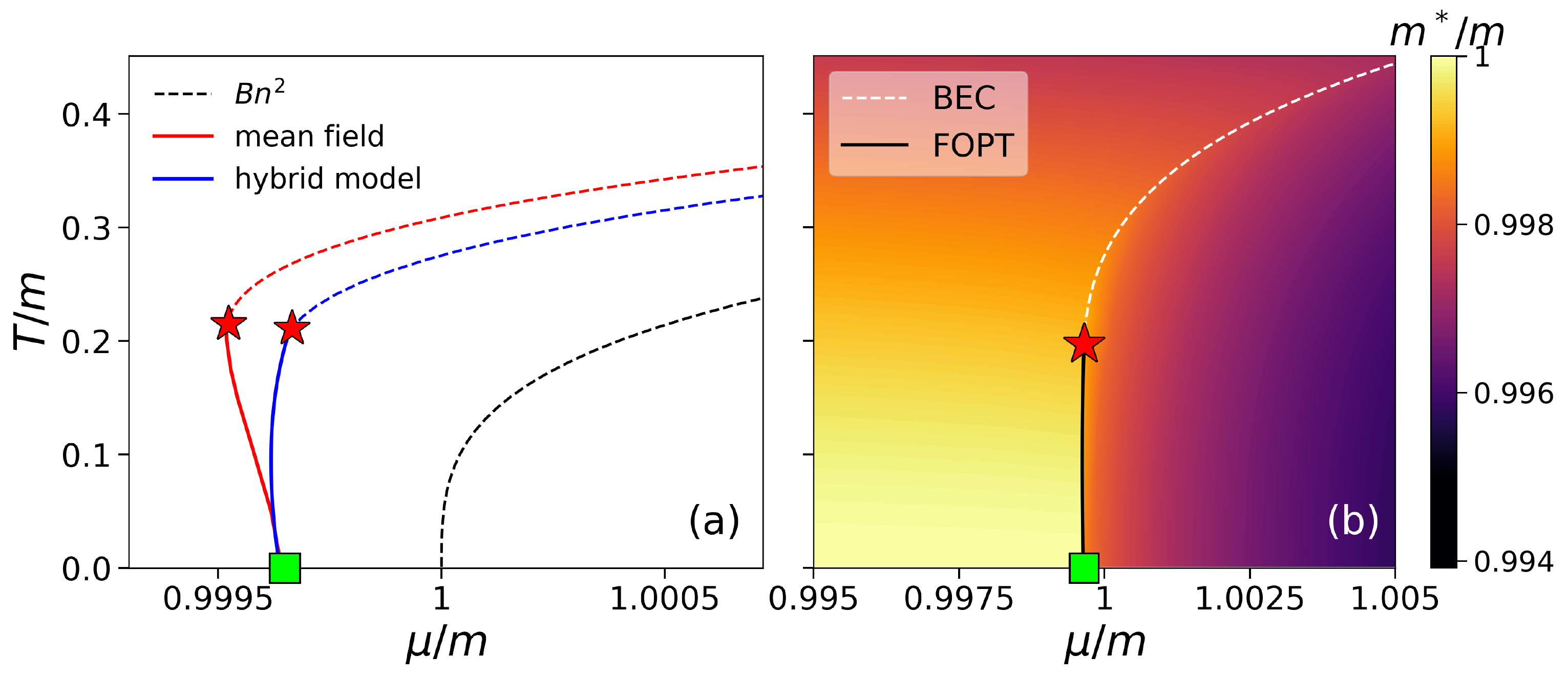}	%
		\caption{\label{fig:6}(a): The FOPT and BEC lines in the mean field and hybrid models on the $(T,\mu)$ plane.  A dotted line presents the pure repulsive interaction ($A=0$) when the FOPT is absent. (b): The $m^*/m$ ratio on the $(\mu,T)$ plane.
		}
    \end{figure*}  

	At the CP,   $dU/dn=0$ and $\omega_Q$ according to (\ref{omegaQ3}) goes to infinity. 
	For $U(n)$ given by Eq.~(\ref{Un}) 
	the condition $dU/dn=0$  gives  
	the  pion number density $n=n_c$ and the chemical potential $\mu=\mu_c$
	at the CP:
	\eq{
		n_c=\frac{A}{2B}~,~~~\mu_c=m-\frac{A^2}{4B}~=~\mu_0- \frac{A^2}{16B}~.
	}
	The critical temperature is calculated numerically as $T_c\cong 0.21~m\cong 28$~MeV. 
	At the CP one also finds $n_+.\gg n_0$ and $n_+\gg n_-$. Therefore, the electric charge density at the CP is approximately 
	\eq{
		n_Q^{c} \cong n_c =  \frac{2}{3} n_0~.
	}
	
	A behavior of $\omega_Q$ on the $(\mu,T)$ plane is shown in Fig.~\ref{fig:5}. 
	Approaching the CP by any path, one observes $\omega_Q\rightarrow \infty$. At $n_Q>0$ this is the only point on the phase diagram with infinite value of the scaled variance $\omega_Q$.  Most interesting regions of the phase diagram are $\mu\cong m$ and $\mu>m$. These are regions of the FOPT  and the BEC.
	For these two phenomena  both the repulsive and attractive interactions between pions play a crucial role.

	\subsection{Hybrid model}	
	Another phenomenological model considered in our paper  is constructed by combining the two frameworks: the mean field $U(n)=Bn^2$ to describe the repulsive interactions (the same as in the mean-field model considered in the previous subsection) and effective pion  mass $m^*(T,\mu)<m$ for attractive ones. It resembles the  Walecka model \cite{Walecka:1974qa}  for the symmetric nuclear matter. A principal difference however is the Bose statistics for interacting pions instead of Fermi statistics for interacting nucleons. There are also some technical  differences. For example, we consider quadratic repulsive mean field $B n^2$ instead of the linear  function of the nucleon number density in Walecka model.  The considered model for interacting pion matter   will be called the hybrid model, and it is defined by the  
	following equations:
	
	\eq{ 
		& p = \sum_{i=+,0,-} p_{\rm id}(T,\mu_i^*;m^*) + \frac{2}{3}B\,n^3 
		-  \frac{(m - m^*)^2}{2A}, \label{phyb} \\
		& n = \sum_{i=+,0,-} n_{\rm id}(T, \mu_i^*;m^*) + n_{\rm BC }^+~,
		\label{nhyb} \\
		& \mu_i^*=\mu_i-Bn^2~,~~~~m^* = m - A n_s, \label{m*}
	}
	where the scalar density
	\eq{n_s=
		\frac{m^*}{2\pi^2}\int_{0}^{\infty}\frac{k^2 d k}{\sqrt{k^2 + m^{*2}}} \,\sum_{i=+,0,-} f_k(T,\mu_i^*;m^*)
		+ n_{\rm BC}^+~.
	}
	\begin{figure*}
		\includegraphics[width=0.95\textwidth]{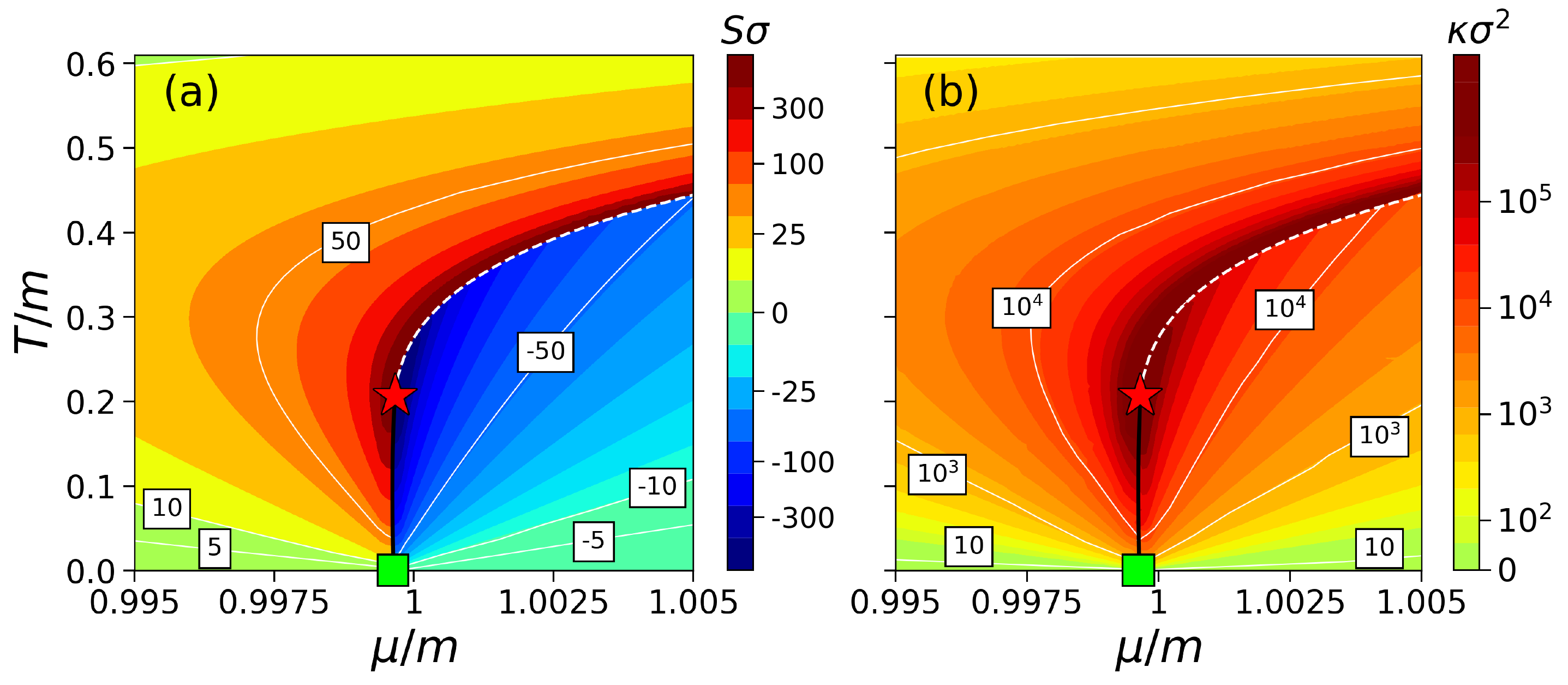}	%
		\caption{\label{fig:7} The skewness (a) and kurtosis (b)  in the hybrid model (see text for details).}
	\end{figure*}
	At $T=0$ one obtains: 
	\eq{
		&p = -~\frac{A}{2}\,\left(n_{\rm BC}^+\right)^2+\frac{2B}{3}\,\left(n_{\rm BC}^+\right)^3~, \label{hyb-1}\\ 
		&\mu^*  =\mu-B (n_{\rm BC}^+)^2~, \label{hyb-2}\\
		&m^*  = m - A n_{\rm BC }
       ^+,~~~\mu^*=m^*~.
\label{hyb-3}
		}
	Equations (\ref{hyb-1})-(\ref{hyb-3}) give the functions $\mu=\mu(n_{BC}^+)$ and $p=p(n_{\rm BC}^+)$ that are identical to the corresponding expressions for the mean-field model with $U(n)$ given by (\ref{Un}) and considered in the previous subsections. Therefore, these two models are identical at $T=0$, and free model parameters of the hybrid model will be taken as in Eq.~(\ref{AB}). 
	At $T>0$ the models are not  identical. However, all conclusions   concerning the FOPT and BEC are the same for the both models. Even more, the numerical values of $T_c$ and $\mu_c$ in the hybrid model are also approximately equal  to those in the mean-field model.
	This is shown in Fig.~\ref{fig:6} (a).
	Because of these reasons we do not present $n_Q(T,\mu)$ and $\omega_Q(T,\mu)$ for the hybrid model. These results are very similar to those in 
	Figs.~\ref{fig:2} and \ref{fig:3} for the mean-field model.
	
	The ratio $m^*/m$ is shown on the $(\mu,T)$ plane for the hybrid model with parameters (\ref{AB}) in Fig.~\ref{fig:6} (b). This ratio is slightly smaller than unity, but rather close to it. 
	This corresponds to the small attractive effects in the hybrid model. In a nonrelativistic approximation one finds $n_s\cong n$ and observes a straightforward  correspondence of the effective mass $m^*$ to the  attractive part of the mean field potential $-An$:
	\eq{
		& \sqrt{(m^*)^2+k^2}\cong m^*+ \frac{k^2}{2m^*} \cong m -An +\frac{k^2}{2m}  \\
		& \frac{(m-m^*)^2}{2A} \cong \frac{1}{2}\, A n^2.
	}
	Nevertheless some differences between these two models can be found.
	For example, the ratio of the pressure function in the hybrid model to pressure of the IdPG at $\mu=0$ is shown as a function of $T$ by a dashed line in Fig.~\ref{fig:2} (b).
	The two models with the same $A$ and $B$ parameters lead to different $p=p(T)$ functions.

	\subsection{Fluctuations of higher orders}\label{FHO}
	Electric (isospin) charge susceptibilities $\chi_j$ in the grand canonical ensemble 
	defined as ($j=1,2,\ldots)$:
	\begin{equation}\label{eq:33}
	\chi_j = \frac{\partial^j (p/T^4)}{\partial (\mu/T)^j}.
	\end{equation}
	Susceptibilities are derivatives of the thermodynamic potential and give, therefore, additional important information of the equation of state. Particularly their values are very sensitive to both the CP and BEC phenomena.  Some ratios of the susceptibilities  \eqref{eq:33} are  well known and used to quantify the fluctuations of conserved charges. Most familiar of these measures are  the scaled variance $\omega$, skewness $S\sigma$, and kurtosis $\kappa\sigma^2$ (see, e.g., Ref. \cite{Karsch_2011}):
	\begin{equation}
	\omega = \frac{\chi_2}{\chi_1}, \quad S\sigma = \frac{\chi_3}{\chi_2}, \quad \kappa\sigma^2 = \frac{\chi_4}{\chi_2}.
	\end{equation}
	
	In this subsection the results for $S\sigma$ and $\kappa\sigma^2$ are presented within the hybrid model with parameters (\ref{AB}).
	These fluctuation measures are presented in  Figs.~\ref{fig:7} (a) and (b), respectively. The results of the mean-field model are essentially the same. 
	At small chemical potential $\mu\ll m$ and not too large $T$  the pion densities are small. In this case 
	both the pion interaction and Bose statistics effects can be neglected. It gives    $S\sigma \cong 1$ and  $\kappa\sigma^2\cong 1$ that corresponds to the ideal classical gas limit.
	Both these measure strongly deviate from these baseline values of the ideal Boltzmann gas at $\mu\cong m$ and $\mu>m$. This is due to a presence of the FOPT and the BEC  
	effects, respectively.

	\paragraph*{Skewness.}
	The skewness $S\sigma$ is presented in Fig.~\ref{fig:7} (a). This measure
	attains both positive and negative values on the $(\mu,T)$ plane. The positive values correspond  to those regions of the phase diagram where
	$n_{\rm BC}^+=0$. The skewness 
	$S\sigma$ has a discontinuity along the BEC line and jumps 
	to negative values in the phase with the  BC $n_{\rm BC}^+>0$.
	At the CP,  $S\sigma$ shows the singular behavior:
	it can go  to both $-\infty$ and  $+\infty$  depending on the path of approaching
	to the CP.  When crossing the FOPT  there is a discontinuity of $S\sigma$ from positive values in the gaseous phase to the negative ones in the liquid phase.

	\paragraph*{Kurtosis.}

	The kurtosis $\kappa\sigma^2$ in the $(\mu,T)$ plane is shown in Fig.~\ref{fig:7} (b).
	The considered models demonstrate only positive values $\kappa\sigma^2>0$ for the whole $(\mu,T)$ plane. 
	This is in contrast to the universal behavior of fluctuations in the Ising model,
	as well as in various phenomenological model calculations
	(see, e.g., Ref.~\cite{Stashko_2021} and references therein), 
	where negative values of $\kappa \sigma^2$ are observed in the so-called analytic crossover region above the critical temperature.
	The large but finite values of the kurtosis are generally obtained in a vicinity of the BEC line with a discontinuity to smaller values on the FOPT and BEC line at increasing $\mu$. 
	The singular behavior with $\kappa\sigma^2\rightarrow \infty$ takes place at the CP.
	The values of the kurtosis remain large with $\kappa\sigma^2 \gg 1$ even far away from the CP. 
	This is due to a large
	sensitivity of the higher order fluctuations to the
	CP.

	\section{Summary}\label{SUM}
	Thermodynamic properties of the interacting pion matter are studied in  the two phenomenological models: the mean-field model with the potential $U(n)$ and the hybrid model. The potential is chosen as a function of the pion number density  $U(n)=-An+Bn^2$, and it includes both the repulsive $Bn^2$ and attractive $-An$ interactions. The hybrid model assumes the same repulsive mean field potential $Bn^2$, while the attractive interactions are described by the effective mass with the excess pressure $-(m-m^*)^2/(2A)$ similar to that in the Walecka model for interacting nucleons \cite{Walecka:1974qa}. Model parameters $A>0$ and $B>0$   %
	are  fixed by fitting the lattice QCD data on the BC pion density at $T=0$ as a function of the chemical potential $\mu$. At zero temperature, the both considered model become identical to each other. Thus, the fitting procedure at $T=0$ leads to the same set (\ref{AB}) of the $A$ and $B$ parameters.
	The two phenomena -- the BEC 
	and the FOPT with the CP 
	-- take place in the pion matter.  
	In spite of some minor differences the two model demonstrate an identical qualitative and very close quantitative  behavior for the thermodynamic functions and electric charge fluctuations in the whole $(\mu,T)$ plane. Note that the qualitative features found in these two models are also in agreement with the results obtained in Ref.~\cite{PhysRevC.103.065201}.   
	
	The interaction parameters $A$ and $B$ (\ref{AB}) found from fitting the lattice data correspond to rather moderate interactions in the pion matter. At $\mu=0$ these interactions are completely unimportant in the pion thermodynamics. It should be emphasized that the mesonic resonances as a part of the pion-pion interactions are not included in our consideration.  
	The residual non-resonance pion-pion interactions are  however crucially important at $\mu\cong m$ and $\mu>m$.

	If $A>0$ and $B>0$ the both models demonstrate the FOPT with a position of the CP at $\mu_c\cong m$ and $T_c\cong 28\,$MeV.
	The BEC line merges to the CP. At $T<T_c$ only the liquid (dense) pion phase includes the BC $n_{\rm BC}^+>0$, while $n_{\rm BC}^+=0$ in the gaseous (rarefied) phase.

	In the ideal pion gas, the scaled variance of electric charge fluctuations   becomes infinite on the BEC line 
	and under this line. In contrast to this ideal gas behavior, a presence of the repulsive interactions makes $\omega_Q$ to be finite and continuous function. The only point of anomalous electric charge fluctuations is the CP. 
	At the CP both the scaled variance  $\omega_Q\rightarrow \infty$ and kurtosis $\kappa\sigma^2\rightarrow \infty$. The skewness $S\sigma$ 
	has a more complicated behavior. It can go to both $+\infty$ and $-\infty$ depending on the way of approaching the CP. A special feature of the considered models is an absence of negative values of the kurtosis. The negative values $\kappa\sigma^2 <0$ are usually happen in a crossover region near the CP. These negative values are absent in the considered models of the pion matter. This is because of the fact that two phenomena -- an  onset of the BEC and the CP -- takes place at the same point.

	Both the FOPT and BEC are mainly defined by  $\pi^+$ mesons. A presence of $\pi^-$ and $\pi^0$ mesons give only moderate numerical corrections and does not change the qualitative properties of the pion matter at $\mu\cong m$ and $\mu>m$.

	 The critical point in the system of interacting pions can be searched in the lQCD. 
The lattice simulations should focus on direct computations of the charge fluctuations measures and include some ``small'' temperatures between 0 and $50$~MeV, i.e., in a vicinity of the hypothetical critical point.

	\begin{acknowledgments}
	The authors thank L. Satarov, H. Stoecker, and V. Vovchenko for fruitful comments and discussions. This work is supported by the National Academy of Sciences of Ukraine, Grant No. 0121U112254. O.S.S. acknowledges the support from National Research Foundation of Ukraine (Project No. 2020.02/0073).
	\end{acknowledgments}
	
	\begin{widetext}
		
		\appendix	
		\section{Derivation of Eq.~(\ref{omegaQ2})}
		In the Appendix we present a proof of Eq.~(\ref{omegaQ2}). 
		The derivative $\partial n_Q/\partial \mu$ can be calculated as
		\eq{ \label{A2}
			T~\frac{\partial n_Q}{\partial\mu} = 
			(n_+\omega_+-\omega_-n_-)\,\frac{ \partial \mu^*}{\partial \mu} +2\omega_-n_-.
		}
		One then finds
		\eq{ \label{A3}
			\frac{\partial \mu^*}{\partial \mu}~=~ 1 ~ -~\frac{1}{T}\,\frac{dU}{dn}\,\left[(n_+\omega_++n_0 \omega_0 + n_-\omega_-) \frac{\partial \mu^*}{\partial \mu}-n_0\omega_0- 2n_-\omega_-\right]~,
		}
		and thus
		\eq{\label{A4}
			\frac{\partial \mu^*}{\partial \mu}~=~1~-~
			\frac{(dU/dn)~(n_+\omega_+ - n_-\omega_-)}
			{T+(dU/(dn) (n_+\omega_++n_0\omega_0+n_-\omega_-)}~.
		}
		Substituting $\partial \mu^*/\partial \mu$ in Eq.~(\ref{A2})  from (\ref{A4}) one obtains Eq.~(\ref{omegaQ2}).
		
	\end{widetext}
	
	\bibliography{references.bib}

\begin{thebibliography}{51}%
\makeatletter
\providecommand \@ifxundefined [1]{%
 \@ifx{#1\undefined}
}%
\providecommand \@ifnum [1]{%
 \ifnum #1\expandafter \@firstoftwo
 \else \expandafter \@secondoftwo
 \fi
}%
\providecommand \@ifx [1]{%
 \ifx #1\expandafter \@firstoftwo
 \else \expandafter \@secondoftwo
 \fi
}%
\providecommand \natexlab [1]{#1}%
\providecommand \enquote  [1]{``#1''}%
\providecommand \bibnamefont  [1]{#1}%
\providecommand \bibfnamefont [1]{#1}%
\providecommand \citenamefont [1]{#1}%
\providecommand \href@noop [0]{\@secondoftwo}%
\providecommand \href [0]{\begingroup \@sanitize@url \@href}%
\providecommand \@href[1]{\@@startlink{#1}\@@href}%
\providecommand \@@href[1]{\endgroup#1\@@endlink}%
\providecommand \@sanitize@url [0]{\catcode `\\12\catcode `\$12\catcode
  `\&12\catcode `\#12\catcode `\^12\catcode `\_12\catcode `\%12\relax}%
\providecommand \@@startlink[1]{}%
\providecommand \@@endlink[0]{}%
\providecommand \url  [0]{\begingroup\@sanitize@url \@url }%
\providecommand \@url [1]{\endgroup\@href {#1}{\urlprefix }}%
\providecommand \urlprefix  [0]{URL }%
\providecommand \Eprint [0]{\href }%
\providecommand \doibase [0]{http://dx.doi.org/}%
\providecommand \selectlanguage [0]{\@gobble}%
\providecommand \bibinfo  [0]{\@secondoftwo}%
\providecommand \bibfield  [0]{\@secondoftwo}%
\providecommand \translation [1]{[#1]}%
\providecommand \BibitemOpen [0]{}%
\providecommand \bibitemStop [0]{}%
\providecommand \bibitemNoStop [0]{.\EOS\space}%
\providecommand \EOS [0]{\spacefactor3000\relax}%
\providecommand \BibitemShut  [1]{\csname bibitem#1\endcsname}%
\let\auto@bib@innerbib\@empty
\bibitem [{\citenamefont {Bose}(1924)}]{Bose1924}%
  \BibitemOpen
  \bibfield  {author} {\bibinfo {author} {\bibfnamefont {S.~N.}\ \bibnamefont
  {Bose}},\ }\href {\doibase 10.1007/BF01327326} {\bibfield  {journal}
  {\bibinfo  {journal} {Zeitschrift f{\"u}r Physik}\ }\textbf {\bibinfo
  {volume} {26}},\ \bibinfo {pages} {178} (\bibinfo {year} {1924})}\BibitemShut
  {NoStop}%
\bibitem [{\citenamefont {Einstein}(1925)}]{Einst1925}%
  \BibitemOpen
  \bibfield  {author} {\bibinfo {author} {\bibfnamefont {A.}~\bibnamefont
  {Einstein}},\ }\href@noop {} {\bibfield  {journal} {\bibinfo  {journal} {Kgl.
  Preuss. Akad. Wiss 1}\ } (\bibinfo {year} {1925})}\BibitemShut {NoStop}%
\bibitem [{\citenamefont {Anderson}\ \emph {et~al.}(1995)\citenamefont
  {Anderson}, \citenamefont {Ensher}, \citenamefont {Matthews}, \citenamefont
  {Wieman},\ and\ \citenamefont {Cornell}}]{Anderson:1995gf}%
  \BibitemOpen
  \bibfield  {author} {\bibinfo {author} {\bibfnamefont {M.~H.}\ \bibnamefont
  {Anderson}}, \bibinfo {author} {\bibfnamefont {J.~R.}\ \bibnamefont
  {Ensher}}, \bibinfo {author} {\bibfnamefont {M.~R.}\ \bibnamefont
  {Matthews}}, \bibinfo {author} {\bibfnamefont {C.~E.}\ \bibnamefont
  {Wieman}}, \ and\ \bibinfo {author} {\bibfnamefont {E.~A.}\ \bibnamefont
  {Cornell}},\ }\href {\doibase 10.1126/science.269.5221.198} {\bibfield
  {journal} {\bibinfo  {journal} {Science}\ }\textbf {\bibinfo {volume}
  {269}},\ \bibinfo {pages} {198} (\bibinfo {year} {1995})}\BibitemShut
  {NoStop}%
\bibitem [{\citenamefont {Davis}\ \emph {et~al.}(1995)\citenamefont {Davis},
  \citenamefont {Mewes}, \citenamefont {Andrews}, \citenamefont {van Druten},
  \citenamefont {Durfee}, \citenamefont {Kurn},\ and\ \citenamefont
  {Ketterle}}]{PhysRevLett.75.3969}%
  \BibitemOpen
  \bibfield  {author} {\bibinfo {author} {\bibfnamefont {K.~B.}\ \bibnamefont
  {Davis}}, \bibinfo {author} {\bibfnamefont {M.~O.}\ \bibnamefont {Mewes}},
  \bibinfo {author} {\bibfnamefont {M.~R.}\ \bibnamefont {Andrews}}, \bibinfo
  {author} {\bibfnamefont {N.~J.}\ \bibnamefont {van Druten}}, \bibinfo
  {author} {\bibfnamefont {D.~S.}\ \bibnamefont {Durfee}}, \bibinfo {author}
  {\bibfnamefont {D.~M.}\ \bibnamefont {Kurn}}, \ and\ \bibinfo {author}
  {\bibfnamefont {W.}~\bibnamefont {Ketterle}},\ }\href {\doibase
  10.1103/PhysRevLett.75.3969} {\bibfield  {journal} {\bibinfo  {journal}
  {Phys. Rev. Lett.}\ }\textbf {\bibinfo {volume} {75}},\ \bibinfo {pages}
  {3969} (\bibinfo {year} {1995})}\BibitemShut {NoStop}%
\bibitem [{\citenamefont {Bradley}\ \emph {et~al.}(1995)\citenamefont
  {Bradley}, \citenamefont {Sackett}, \citenamefont {Tollett},\ and\
  \citenamefont {Hulet}}]{PhysRevLett.75.1687}%
  \BibitemOpen
  \bibfield  {author} {\bibinfo {author} {\bibfnamefont {C.~C.}\ \bibnamefont
  {Bradley}}, \bibinfo {author} {\bibfnamefont {C.~A.}\ \bibnamefont
  {Sackett}}, \bibinfo {author} {\bibfnamefont {J.~J.}\ \bibnamefont
  {Tollett}}, \ and\ \bibinfo {author} {\bibfnamefont {R.~G.}\ \bibnamefont
  {Hulet}},\ }\href {\doibase 10.1103/PhysRevLett.75.1687} {\bibfield
  {journal} {\bibinfo  {journal} {Phys. Rev. Lett.}\ }\textbf {\bibinfo
  {volume} {75}},\ \bibinfo {pages} {1687} (\bibinfo {year}
  {1995})}\BibitemShut {NoStop}%
\bibitem [{\citenamefont {Dalfovo}\ \emph {et~al.}(1999)\citenamefont
  {Dalfovo}, \citenamefont {Giorgini}, \citenamefont {Pitaevskii},\ and\
  \citenamefont {Stringari}}]{RevModPhys.71.463}%
  \BibitemOpen
  \bibfield  {author} {\bibinfo {author} {\bibfnamefont {F.}~\bibnamefont
  {Dalfovo}}, \bibinfo {author} {\bibfnamefont {S.}~\bibnamefont {Giorgini}},
  \bibinfo {author} {\bibfnamefont {L.~P.}\ \bibnamefont {Pitaevskii}}, \ and\
  \bibinfo {author} {\bibfnamefont {S.}~\bibnamefont {Stringari}},\ }\href
  {\doibase 10.1103/RevModPhys.71.463} {\bibfield  {journal} {\bibinfo
  {journal} {Rev. Mod. Phys.}\ }\textbf {\bibinfo {volume} {71}},\ \bibinfo
  {pages} {463} (\bibinfo {year} {1999})}\BibitemShut {NoStop}%
\bibitem [{\citenamefont {Satarov}\ \emph {et~al.}(2017)\citenamefont
  {Satarov}, \citenamefont {Gorenstein}, \citenamefont {Motornenko},
  \citenamefont {Vovchenko}, \citenamefont {Mishustin},\ and\ \citenamefont
  {Stoecker}}]{Satarov_2017}%
  \BibitemOpen
  \bibfield  {author} {\bibinfo {author} {\bibfnamefont {L.~M.}\ \bibnamefont
  {Satarov}}, \bibinfo {author} {\bibfnamefont {M.~I.}\ \bibnamefont
  {Gorenstein}}, \bibinfo {author} {\bibfnamefont {A.}~\bibnamefont
  {Motornenko}}, \bibinfo {author} {\bibfnamefont {V.}~\bibnamefont
  {Vovchenko}}, \bibinfo {author} {\bibfnamefont {I.~N.}\ \bibnamefont
  {Mishustin}}, \ and\ \bibinfo {author} {\bibfnamefont {H.}~\bibnamefont
  {Stoecker}},\ }\href {\doibase 10.1088/1361-6471/aa8c5d} {\bibfield
  {journal} {\bibinfo  {journal} {Journal of Physics G: Nuclear and Particle
  Physics}\ }\textbf {\bibinfo {volume} {44}},\ \bibinfo {pages} {125102}
  (\bibinfo {year} {2017})}\BibitemShut {NoStop}%
\bibitem [{\citenamefont {Begun}\ and\ \citenamefont
  {Gorenstein}(2007)}]{Begun:2006gj}%
  \BibitemOpen
  \bibfield  {author} {\bibinfo {author} {\bibfnamefont {V.~V.}\ \bibnamefont
  {Begun}}\ and\ \bibinfo {author} {\bibfnamefont {M.~I.}\ \bibnamefont
  {Gorenstein}},\ }\href {\doibase 10.1016/j.physletb.2007.07.059} {\bibfield
  {journal} {\bibinfo  {journal} {Phys. Lett.}\ }\textbf {\bibinfo {volume}
  {B653}},\ \bibinfo {pages} {190} (\bibinfo {year} {2007})},\ \Eprint
  {http://arxiv.org/abs/hep-ph/0611043} {arXiv:hep-ph/0611043 [hep-ph]}
  \BibitemShut {NoStop}%
\bibitem [{\citenamefont {Begun}\ and\ \citenamefont
  {Gorenstein}(2008)}]{Begun:2008hq}%
  \BibitemOpen
  \bibfield  {author} {\bibinfo {author} {\bibfnamefont {V.~V.}\ \bibnamefont
  {Begun}}\ and\ \bibinfo {author} {\bibfnamefont {M.~I.}\ \bibnamefont
  {Gorenstein}},\ }\href {\doibase 10.1103/PhysRevC.77.064903} {\bibfield
  {journal} {\bibinfo  {journal} {Phys. Rev.}\ }\textbf {\bibinfo {volume}
  {C77}},\ \bibinfo {pages} {064903} (\bibinfo {year} {2008})},\ \Eprint
  {http://arxiv.org/abs/0802.3349} {arXiv:0802.3349 [hep-ph]} \BibitemShut
  {NoStop}%
\bibitem [{\citenamefont {Strinati}\ \emph {et~al.}(2018)\citenamefont
  {Strinati}, \citenamefont {Pieri}, \citenamefont {Röpke}, \citenamefont
  {Schuck},\ and\ \citenamefont {Urban}}]{Strinati_2018}%
  \BibitemOpen
  \bibfield  {author} {\bibinfo {author} {\bibfnamefont {G.~C.}\ \bibnamefont
  {Strinati}}, \bibinfo {author} {\bibfnamefont {P.}~\bibnamefont {Pieri}},
  \bibinfo {author} {\bibfnamefont {G.}~\bibnamefont {Röpke}}, \bibinfo
  {author} {\bibfnamefont {P.}~\bibnamefont {Schuck}}, \ and\ \bibinfo {author}
  {\bibfnamefont {M.}~\bibnamefont {Urban}},\ }\href {\doibase
  10.1016/j.physrep.2018.02.004} {\bibfield  {journal} {\bibinfo  {journal}
  {Physics Reports}\ }\textbf {\bibinfo {volume} {738}},\ \bibinfo {pages}
  {1–76} (\bibinfo {year} {2018})}\BibitemShut {NoStop}%
\bibitem [{\citenamefont {Nozieres}\ and\ \citenamefont
  {Schmitt-Rink}(1985)}]{Nozieres:1985zz}%
  \BibitemOpen
  \bibfield  {author} {\bibinfo {author} {\bibfnamefont {P.}~\bibnamefont
  {Nozieres}}\ and\ \bibinfo {author} {\bibfnamefont {S.}~\bibnamefont
  {Schmitt-Rink}},\ }\href {\doibase 10.1007/BF00683774} {\bibfield  {journal}
  {\bibinfo  {journal} {J. Low. Temp. Phys.}\ }\textbf {\bibinfo {volume}
  {59}},\ \bibinfo {pages} {195} (\bibinfo {year} {1985})}\BibitemShut
  {NoStop}%
\bibitem [{\citenamefont {Funaki}\ \emph {et~al.}(2008)\citenamefont {Funaki},
  \citenamefont {Yamada}, \citenamefont {Horiuchi}, \citenamefont {R\"opke},
  \citenamefont {Schuck},\ and\ \citenamefont
  {Tohsaki}}]{PhysRevLett.101.082502}%
  \BibitemOpen
  \bibfield  {author} {\bibinfo {author} {\bibfnamefont {Y.}~\bibnamefont
  {Funaki}}, \bibinfo {author} {\bibfnamefont {T.}~\bibnamefont {Yamada}},
  \bibinfo {author} {\bibfnamefont {H.}~\bibnamefont {Horiuchi}}, \bibinfo
  {author} {\bibfnamefont {G.}~\bibnamefont {R\"opke}}, \bibinfo {author}
  {\bibfnamefont {P.}~\bibnamefont {Schuck}}, \ and\ \bibinfo {author}
  {\bibfnamefont {A.}~\bibnamefont {Tohsaki}},\ }\href {\doibase
  10.1103/PhysRevLett.101.082502} {\bibfield  {journal} {\bibinfo  {journal}
  {Phys. Rev. Lett.}\ }\textbf {\bibinfo {volume} {101}},\ \bibinfo {pages}
  {082502} (\bibinfo {year} {2008})}\BibitemShut {NoStop}%
\bibitem [{\citenamefont {Chavanis}\ and\ \citenamefont
  {Harko}(2012)}]{Chavanis:2011cz}%
  \BibitemOpen
  \bibfield  {author} {\bibinfo {author} {\bibfnamefont {P.-H.}\ \bibnamefont
  {Chavanis}}\ and\ \bibinfo {author} {\bibfnamefont {T.}~\bibnamefont
  {Harko}},\ }\href {\doibase 10.1103/PhysRevD.86.064011} {\bibfield  {journal}
  {\bibinfo  {journal} {Phys. Rev.}\ }\textbf {\bibinfo {volume} {D86}},\
  \bibinfo {pages} {064011} (\bibinfo {year} {2012})},\ \Eprint
  {http://arxiv.org/abs/1108.3986} {arXiv:1108.3986 [astro-ph.SR]} \BibitemShut
  {NoStop}%
\bibitem [{\citenamefont {Mishustin}\ \emph {et~al.}(2019)\citenamefont
  {Mishustin}, \citenamefont {Anchishkin}, \citenamefont {Satarov},
  \citenamefont {Stashko},\ and\ \citenamefont {Stoecker}}]{Mishustin_2019}%
  \BibitemOpen
  \bibfield  {author} {\bibinfo {author} {\bibfnamefont {I.}~\bibnamefont
  {Mishustin}}, \bibinfo {author} {\bibfnamefont {D.}~\bibnamefont
  {Anchishkin}}, \bibinfo {author} {\bibfnamefont {L.}~\bibnamefont {Satarov}},
  \bibinfo {author} {\bibfnamefont {O.}~\bibnamefont {Stashko}}, \ and\
  \bibinfo {author} {\bibfnamefont {H.}~\bibnamefont {Stoecker}},\ }\href
  {\doibase 10.1103/PhysRevC.100.022201} {\bibfield  {journal} {\bibinfo
  {journal} {Phys.\ Rev.\ C}\ }\textbf {\bibinfo {volume} {100}},\ \bibinfo
  {pages} {022201} (\bibinfo {year} {2019})},\ \Eprint
  {http://arxiv.org/abs/1905.09567} {arXiv:1905.09567 [nucl-th]} \BibitemShut
  {NoStop}%
\bibitem [{\citenamefont {Padilla}\ \emph {et~al.}(2019)\citenamefont
  {Padilla}, \citenamefont {V{\'{a}}zquez}, \citenamefont {Matos},\ and\
  \citenamefont {Germ{\'{a}}n}}]{Padilla_2019}%
  \BibitemOpen
  \bibfield  {author} {\bibinfo {author} {\bibfnamefont {L.~E.}\ \bibnamefont
  {Padilla}}, \bibinfo {author} {\bibfnamefont {J.~A.}\ \bibnamefont
  {V{\'{a}}zquez}}, \bibinfo {author} {\bibfnamefont {T.}~\bibnamefont
  {Matos}}, \ and\ \bibinfo {author} {\bibfnamefont {G.}~\bibnamefont
  {Germ{\'{a}}n}},\ }\href {\doibase 10.1088/1475-7516/2019/05/056} {\bibfield
  {journal} {\bibinfo  {journal} {Journal of Cosmology and Astroparticle
  Physics}\ }\textbf {\bibinfo {volume} {2019}},\ \bibinfo {pages} {056}
  (\bibinfo {year} {2019})}\BibitemShut {NoStop}%
\bibitem [{\citenamefont {Vovchenko}\ \emph {et~al.}(2021)\citenamefont
  {Vovchenko}, \citenamefont {Brandt}, \citenamefont {Cuteri}, \citenamefont
  {Endrődi}, \citenamefont {Hajkarim},\ and\ \citenamefont
  {Schaffner-Bielich}}]{Vovchenko_2021}%
  \BibitemOpen
  \bibfield  {author} {\bibinfo {author} {\bibfnamefont {V.}~\bibnamefont
  {Vovchenko}}, \bibinfo {author} {\bibfnamefont {B.~B.}\ \bibnamefont
  {Brandt}}, \bibinfo {author} {\bibfnamefont {F.}~\bibnamefont {Cuteri}},
  \bibinfo {author} {\bibfnamefont {G.}~\bibnamefont {Endrődi}}, \bibinfo
  {author} {\bibfnamefont {F.}~\bibnamefont {Hajkarim}}, \ and\ \bibinfo
  {author} {\bibfnamefont {J.}~\bibnamefont {Schaffner-Bielich}},\ }\href
  {\doibase 10.1103/physrevlett.126.012701} {\bibfield  {journal} {\bibinfo
  {journal} {Physical Review Letters}\ }\textbf {\bibinfo {volume} {126}}
  (\bibinfo {year} {2021}),\ 10.1103/physrevlett.126.012701}\BibitemShut
  {NoStop}%
\bibitem [{\citenamefont {Brandt}\ \emph
  {et~al.}(2018{\natexlab{a}})\citenamefont {Brandt}, \citenamefont {Endrodi},\
  and\ \citenamefont {Schmalzbauer}}]{Brandt_2018}%
  \BibitemOpen
  \bibfield  {author} {\bibinfo {author} {\bibfnamefont {B.~B.}\ \bibnamefont
  {Brandt}}, \bibinfo {author} {\bibfnamefont {G.}~\bibnamefont {Endrodi}}, \
  and\ \bibinfo {author} {\bibfnamefont {S.}~\bibnamefont {Schmalzbauer}},\
  }\href {\doibase 10.1051/epjconf/201817507020} {\bibfield  {journal}
  {\bibinfo  {journal} {EPJ Web Conf.}\ }\textbf {\bibinfo {volume} {175}},\
  \bibinfo {pages} {07020} (\bibinfo {year} {2018}{\natexlab{a}})},\ \Eprint
  {http://arxiv.org/abs/1709.10487} {arXiv:1709.10487 [hep-lat]} \BibitemShut
  {NoStop}%
\bibitem [{\citenamefont {Mannarelli}(2019)}]{Mannarelli_2019}%
  \BibitemOpen
  \bibfield  {author} {\bibinfo {author} {\bibfnamefont {M.}~\bibnamefont
  {Mannarelli}},\ }\href {\doibase 10.3390/particles2030025} {\bibfield
  {journal} {\bibinfo  {journal} {Particles}\ }\textbf {\bibinfo {volume}
  {2}},\ \bibinfo {pages} {411–443} (\bibinfo {year} {2019})}\BibitemShut
  {NoStop}%
\bibitem [{\citenamefont {{Andersen}}\ and\ \citenamefont
  {{Kneschke}}(2018)}]{andersen2018boseeinstein}%
  \BibitemOpen
  \bibfield  {author} {\bibinfo {author} {\bibfnamefont {J.~O.}\ \bibnamefont
  {{Andersen}}}\ and\ \bibinfo {author} {\bibfnamefont {P.}~\bibnamefont
  {{Kneschke}}},\ }\href@noop {} {\bibfield  {journal} {\bibinfo  {journal}
  {arXiv e-prints}\ ,\ \bibinfo {eid} {arXiv:1807.08951}} (\bibinfo {year}
  {2018})},\ \Eprint {http://arxiv.org/abs/1807.08951} {arXiv:1807.08951
  [hep-ph]} \BibitemShut {NoStop}%
\bibitem [{\citenamefont {Begun}\ and\ \citenamefont
  {Florkowski}(2015)}]{Begun_2015}%
  \BibitemOpen
  \bibfield  {author} {\bibinfo {author} {\bibfnamefont {V.}~\bibnamefont
  {Begun}}\ and\ \bibinfo {author} {\bibfnamefont {W.}~\bibnamefont
  {Florkowski}},\ }\href {\doibase 10.1103/physrevc.91.054909} {\bibfield
  {journal} {\bibinfo  {journal} {Physical Review C}\ }\textbf {\bibinfo
  {volume} {91}} (\bibinfo {year} {2015}),\
  10.1103/physrevc.91.054909}\BibitemShut {NoStop}%
\bibitem [{\citenamefont {Son}\ and\ \citenamefont
  {Stephanov}(2001)}]{Son:2000xc}%
  \BibitemOpen
  \bibfield  {author} {\bibinfo {author} {\bibfnamefont {D.~T.}\ \bibnamefont
  {Son}}\ and\ \bibinfo {author} {\bibfnamefont {M.~A.}\ \bibnamefont
  {Stephanov}},\ }\href {\doibase 10.1103/PhysRevLett.86.592} {\bibfield
  {journal} {\bibinfo  {journal} {Phys. Rev. Lett.}\ }\textbf {\bibinfo
  {volume} {86}},\ \bibinfo {pages} {592} (\bibinfo {year} {2001})},\ \Eprint
  {http://arxiv.org/abs/hep-ph/0005225} {arXiv:hep-ph/0005225 [hep-ph]}
  \BibitemShut {NoStop}%
\bibitem [{\citenamefont {Abuki}\ \emph {et~al.}(2009)\citenamefont {Abuki},
  \citenamefont {Brauner},\ and\ \citenamefont {Warringa}}]{Abuki:2009hx}%
  \BibitemOpen
  \bibfield  {author} {\bibinfo {author} {\bibfnamefont {H.}~\bibnamefont
  {Abuki}}, \bibinfo {author} {\bibfnamefont {T.}~\bibnamefont {Brauner}}, \
  and\ \bibinfo {author} {\bibfnamefont {H.~J.}\ \bibnamefont {Warringa}},\
  }\href {\doibase 10.1140/epjc/s10052-009-1121-0} {\bibfield  {journal}
  {\bibinfo  {journal} {Eur.\ Phys.\ J.\ C}\ }\textbf {\bibinfo {volume}
  {64}},\ \bibinfo {pages} {123} (\bibinfo {year} {2009})},\ \Eprint
  {http://arxiv.org/abs/0901.2477} {arXiv:0901.2477 [hep-ph]} \BibitemShut
  {NoStop}%
\bibitem [{\citenamefont {Brandt}\ \emph
  {et~al.}(2018{\natexlab{b}})\citenamefont {Brandt}, \citenamefont {Endrodi},\
  and\ \citenamefont {Schmalzbauer}}]{Brandt:2017oyy}%
  \BibitemOpen
  \bibfield  {author} {\bibinfo {author} {\bibfnamefont {B.~B.}\ \bibnamefont
  {Brandt}}, \bibinfo {author} {\bibfnamefont {G.}~\bibnamefont {Endrodi}}, \
  and\ \bibinfo {author} {\bibfnamefont {S.}~\bibnamefont {Schmalzbauer}},\
  }\href {\doibase 10.1103/PhysRevD.97.054514} {\bibfield  {journal} {\bibinfo
  {journal} {Phys. Rev.}\ }\textbf {\bibinfo {volume} {D97}},\ \bibinfo {pages}
  {054514} (\bibinfo {year} {2018}{\natexlab{b}})},\ \Eprint
  {http://arxiv.org/abs/1712.08190} {arXiv:1712.08190 [hep-lat]} \BibitemShut
  {NoStop}%
\bibitem [{\citenamefont {Brandt}\ \emph
  {et~al.}(2018{\natexlab{c}})\citenamefont {Brandt}, \citenamefont {Endrodi},
  \citenamefont {Fraga}, \citenamefont {Hippert}, \citenamefont
  {Schaffner-Bielich},\ and\ \citenamefont {Schmalzbauer}}]{Brandt:2018bwq}%
  \BibitemOpen
  \bibfield  {author} {\bibinfo {author} {\bibfnamefont {B.~B.}\ \bibnamefont
  {Brandt}}, \bibinfo {author} {\bibfnamefont {G.}~\bibnamefont {Endrodi}},
  \bibinfo {author} {\bibfnamefont {E.~S.}\ \bibnamefont {Fraga}}, \bibinfo
  {author} {\bibfnamefont {M.}~\bibnamefont {Hippert}}, \bibinfo {author}
  {\bibfnamefont {J.}~\bibnamefont {Schaffner-Bielich}}, \ and\ \bibinfo
  {author} {\bibfnamefont {S.}~\bibnamefont {Schmalzbauer}},\ }\href {\doibase
  10.1103/PhysRevD.98.094510} {\bibfield  {journal} {\bibinfo  {journal} {Phys.
  Rev.}\ }\textbf {\bibinfo {volume} {D98}},\ \bibinfo {pages} {094510}
  (\bibinfo {year} {2018}{\natexlab{c}})},\ \Eprint
  {http://arxiv.org/abs/1802.06685} {arXiv:1802.06685 [hep-ph]} \BibitemShut
  {NoStop}%
\bibitem [{\citenamefont {Adhikari}\ and\ \citenamefont
  {Andersen}(2020)}]{Adhikari:2019zaj}%
  \BibitemOpen
  \bibfield  {author} {\bibinfo {author} {\bibfnamefont {P.}~\bibnamefont
  {Adhikari}}\ and\ \bibinfo {author} {\bibfnamefont {J.~O.}\ \bibnamefont
  {Andersen}},\ }\href {\doibase 10.1016/j.physletb.2020.135352} {\bibfield
  {journal} {\bibinfo  {journal} {Phys. Lett. B}\ }\textbf {\bibinfo {volume}
  {804}},\ \bibinfo {pages} {135352} (\bibinfo {year} {2020})},\ \Eprint
  {http://arxiv.org/abs/1909.01131} {arXiv:1909.01131 [hep-ph]} \BibitemShut
  {NoStop}%
\bibitem [{\citenamefont {Adhikari}\ \emph {et~al.}(2020)\citenamefont
  {Adhikari}, \citenamefont {Andersen},\ and\ \citenamefont
  {Mojahed}}]{Adhikari:2020kdn}%
  \BibitemOpen
  \bibfield  {author} {\bibinfo {author} {\bibfnamefont {P.}~\bibnamefont
  {Adhikari}}, \bibinfo {author} {\bibfnamefont {J.~O.}\ \bibnamefont
  {Andersen}}, \ and\ \bibinfo {author} {\bibfnamefont {M.~A.}\ \bibnamefont
  {Mojahed}},\ }\href@noop {} {\  (\bibinfo {year} {2020})},\ \Eprint
  {http://arxiv.org/abs/2010.13655} {arXiv:2010.13655 [hep-ph]} \BibitemShut
  {NoStop}%
\bibitem [{\citenamefont {Andersen}(2007)}]{Andersen:2006ys}%
  \BibitemOpen
  \bibfield  {author} {\bibinfo {author} {\bibfnamefont {J.~O.}\ \bibnamefont
  {Andersen}},\ }\href {\doibase 10.1103/physrevd.75.065011} {\bibfield
  {journal} {\bibinfo  {journal} {Physical Review D}\ }\textbf {\bibinfo
  {volume} {75}} (\bibinfo {year} {2007}),\
  10.1103/physrevd.75.065011}\BibitemShut {NoStop}%
\bibitem [{\citenamefont {Andersen}\ and\ \citenamefont
  {Brauner}(2008)}]{Andersen:2008qk}%
  \BibitemOpen
  \bibfield  {author} {\bibinfo {author} {\bibfnamefont {J.~O.}\ \bibnamefont
  {Andersen}}\ and\ \bibinfo {author} {\bibfnamefont {T.}~\bibnamefont
  {Brauner}},\ }\href {\doibase 10.1103/PhysRevD.78.014030} {\bibfield
  {journal} {\bibinfo  {journal} {Phys. Rev. D}\ }\textbf {\bibinfo {volume}
  {78}},\ \bibinfo {pages} {014030} (\bibinfo {year} {2008})},\ \Eprint
  {http://arxiv.org/abs/0804.4604} {arXiv:0804.4604 [hep-ph]} \BibitemShut
  {NoStop}%
\bibitem [{\citenamefont {He}\ \emph {et~al.}(2005)\citenamefont {He},
  \citenamefont {Jin},\ and\ \citenamefont {Zhuang}}]{PhysRevD.71.116001}%
  \BibitemOpen
  \bibfield  {author} {\bibinfo {author} {\bibfnamefont {L.}~\bibnamefont
  {He}}, \bibinfo {author} {\bibfnamefont {M.}~\bibnamefont {Jin}}, \ and\
  \bibinfo {author} {\bibfnamefont {P.}~\bibnamefont {Zhuang}},\ }\href
  {\doibase 10.1103/PhysRevD.71.116001} {\bibfield  {journal} {\bibinfo
  {journal} {Phys. Rev. D}\ }\textbf {\bibinfo {volume} {71}},\ \bibinfo
  {pages} {116001} (\bibinfo {year} {2005})}\BibitemShut {NoStop}%
\bibitem [{\citenamefont {Adhikari}\ \emph {et~al.}(2018)\citenamefont
  {Adhikari}, \citenamefont {Andersen},\ and\ \citenamefont
  {Kneschke}}]{PhysRevD.98.074016}%
  \BibitemOpen
  \bibfield  {author} {\bibinfo {author} {\bibfnamefont {P.}~\bibnamefont
  {Adhikari}}, \bibinfo {author} {\bibfnamefont {J.~O.}\ \bibnamefont
  {Andersen}}, \ and\ \bibinfo {author} {\bibfnamefont {P.}~\bibnamefont
  {Kneschke}},\ }\href {\doibase 10.1103/PhysRevD.98.074016} {\bibfield
  {journal} {\bibinfo  {journal} {Phys. Rev. D}\ }\textbf {\bibinfo {volume}
  {98}},\ \bibinfo {pages} {074016} (\bibinfo {year} {2018})}\BibitemShut
  {NoStop}%
\bibitem [{\citenamefont {Folkestad}\ and\ \citenamefont
  {Andersen}(2019)}]{Folkestad_2019}%
  \BibitemOpen
  \bibfield  {author} {\bibinfo {author} {\bibfnamefont {Ã.}~\bibnamefont
  {Folkestad}}\ and\ \bibinfo {author} {\bibfnamefont {J.~O.}\ \bibnamefont
  {Andersen}},\ }\href {\doibase 10.1103/physrevd.99.054006} {\bibfield
  {journal} {\bibinfo  {journal} {Physical Review D}\ }\textbf {\bibinfo
  {volume} {99}} (\bibinfo {year} {2019}),\
  10.1103/physrevd.99.054006}\BibitemShut {NoStop}%
\bibitem [{\citenamefont {Kamikado}\ \emph {et~al.}(2013)\citenamefont
  {Kamikado}, \citenamefont {Strodthoff}, \citenamefont {{von Smekal}},\ and\
  \citenamefont {Wambach}}]{KAMIKADO20131044}%
  \BibitemOpen
  \bibfield  {author} {\bibinfo {author} {\bibfnamefont {K.}~\bibnamefont
  {Kamikado}}, \bibinfo {author} {\bibfnamefont {N.}~\bibnamefont
  {Strodthoff}}, \bibinfo {author} {\bibfnamefont {L.}~\bibnamefont {{von
  Smekal}}}, \ and\ \bibinfo {author} {\bibfnamefont {J.}~\bibnamefont
  {Wambach}},\ }\href {\doibase https://doi.org/10.1016/j.physletb.2012.11.055}
  {\bibfield  {journal} {\bibinfo  {journal} {Physics Letters B}\ }\textbf
  {\bibinfo {volume} {718}},\ \bibinfo {pages} {1044} (\bibinfo {year}
  {2013})}\BibitemShut {NoStop}%
\bibitem [{\citenamefont {Svanes}\ and\ \citenamefont
  {Andersen}(2011)}]{SVANES201116}%
  \BibitemOpen
  \bibfield  {author} {\bibinfo {author} {\bibfnamefont {E.~E.}\ \bibnamefont
  {Svanes}}\ and\ \bibinfo {author} {\bibfnamefont {J.~O.}\ \bibnamefont
  {Andersen}},\ }\href {\doibase
  https://doi.org/10.1016/j.nuclphysa.2011.03.007} {\bibfield  {journal}
  {\bibinfo  {journal} {Nuclear Physics A}\ }\textbf {\bibinfo {volume}
  {857}},\ \bibinfo {pages} {16} (\bibinfo {year} {2011})}\BibitemShut
  {NoStop}%
\bibitem [{\citenamefont {Andersen}\ \emph {et~al.}(2016)\citenamefont
  {Andersen}, \citenamefont {Haque}, \citenamefont {Mustafa},\ and\
  \citenamefont {Strickland}}]{PhysRevD.93.054045}%
  \BibitemOpen
  \bibfield  {author} {\bibinfo {author} {\bibfnamefont {J.~O.}\ \bibnamefont
  {Andersen}}, \bibinfo {author} {\bibfnamefont {N.}~\bibnamefont {Haque}},
  \bibinfo {author} {\bibfnamefont {M.~G.}\ \bibnamefont {Mustafa}}, \ and\
  \bibinfo {author} {\bibfnamefont {M.}~\bibnamefont {Strickland}},\ }\href
  {\doibase 10.1103/physrevd.93.054045} {\bibfield  {journal} {\bibinfo
  {journal} {Physical Review D}\ }\textbf {\bibinfo {volume} {93}} (\bibinfo
  {year} {2016}),\ 10.1103/physrevd.93.054045}\BibitemShut {NoStop}%
\bibitem [{\citenamefont {Stashko}\ \emph
  {et~al.}(2021{\natexlab{a}})\citenamefont {Stashko}, \citenamefont
  {Anchishkin}, \citenamefont {Savchuk},\ and\ \citenamefont
  {Gorenstein}}]{Stashko_2021}%
  \BibitemOpen
  \bibfield  {author} {\bibinfo {author} {\bibfnamefont {O.~S.}\ \bibnamefont
  {Stashko}}, \bibinfo {author} {\bibfnamefont {D.~V.}\ \bibnamefont
  {Anchishkin}}, \bibinfo {author} {\bibfnamefont {O.~V.}\ \bibnamefont
  {Savchuk}}, \ and\ \bibinfo {author} {\bibfnamefont {M.~I.}\ \bibnamefont
  {Gorenstein}},\ }\href {\doibase 10.1088/1361-6471/abd5a5} {\bibfield
  {journal} {\bibinfo  {journal} {Journal of Physics G: Nuclear and Particle
  Physics}\ }\textbf {\bibinfo {volume} {48}},\ \bibinfo {pages} {055106}
  (\bibinfo {year} {2021}{\natexlab{a}})}\BibitemShut {NoStop}%
\bibitem [{\citenamefont {Dashen}\ \emph {et~al.}(1969)\citenamefont {Dashen},
  \citenamefont {Ma},\ and\ \citenamefont {Bernstein}}]{Dashen:1969ep}%
  \BibitemOpen
  \bibfield  {author} {\bibinfo {author} {\bibfnamefont {R.}~\bibnamefont
  {Dashen}}, \bibinfo {author} {\bibfnamefont {S.-K.}\ \bibnamefont {Ma}}, \
  and\ \bibinfo {author} {\bibfnamefont {H.~J.}\ \bibnamefont {Bernstein}},\
  }\href {\doibase 10.1103/PhysRev.187.345} {\bibfield  {journal} {\bibinfo
  {journal} {Phys. Rev.}\ }\textbf {\bibinfo {volume} {187}},\ \bibinfo {pages}
  {345} (\bibinfo {year} {1969})}\BibitemShut {NoStop}%
\bibitem [{\citenamefont {Venugopalan}\ and\ \citenamefont
  {Prakash}(1992)}]{Venugopalan:1992hy}%
  \BibitemOpen
  \bibfield  {author} {\bibinfo {author} {\bibfnamefont {R.}~\bibnamefont
  {Venugopalan}}\ and\ \bibinfo {author} {\bibfnamefont {M.}~\bibnamefont
  {Prakash}},\ }\href {\doibase 10.1016/0375-9474(92)90005-5} {\bibfield
  {journal} {\bibinfo  {journal} {Nucl. Phys. A}\ }\textbf {\bibinfo {volume}
  {546}},\ \bibinfo {pages} {718} (\bibinfo {year} {1992})}\BibitemShut
  {NoStop}%
\bibitem [{\citenamefont {Broniowski}\ \emph {et~al.}(2015)\citenamefont
  {Broniowski}, \citenamefont {Giacosa},\ and\ \citenamefont
  {Begun}}]{Broniowski:2015oha}%
  \BibitemOpen
  \bibfield  {author} {\bibinfo {author} {\bibfnamefont {W.}~\bibnamefont
  {Broniowski}}, \bibinfo {author} {\bibfnamefont {F.}~\bibnamefont {Giacosa}},
  \ and\ \bibinfo {author} {\bibfnamefont {V.}~\bibnamefont {Begun}},\ }\href
  {\doibase 10.1103/PhysRevC.92.034905} {\bibfield  {journal} {\bibinfo
  {journal} {Phys. Rev. C}\ }\textbf {\bibinfo {volume} {92}},\ \bibinfo
  {pages} {034905} (\bibinfo {year} {2015})},\ \Eprint
  {http://arxiv.org/abs/1506.01260} {arXiv:1506.01260 [nucl-th]} \BibitemShut
  {NoStop}%
\bibitem [{\citenamefont {Savchuk}\ \emph {et~al.}(2020)\citenamefont
  {Savchuk}, \citenamefont {Bondar}, \citenamefont {Stashko}, \citenamefont
  {Poberezhnyuk}, \citenamefont {Vovchenko}, \citenamefont {Gorenstein},\ and\
  \citenamefont {Stoecker}}]{Savch2020}%
  \BibitemOpen
  \bibfield  {author} {\bibinfo {author} {\bibfnamefont {O.}~\bibnamefont
  {Savchuk}}, \bibinfo {author} {\bibfnamefont {Y.}~\bibnamefont {Bondar}},
  \bibinfo {author} {\bibfnamefont {O.}~\bibnamefont {Stashko}}, \bibinfo
  {author} {\bibfnamefont {R.~V.}\ \bibnamefont {Poberezhnyuk}}, \bibinfo
  {author} {\bibfnamefont {V.}~\bibnamefont {Vovchenko}}, \bibinfo {author}
  {\bibfnamefont {M.~I.}\ \bibnamefont {Gorenstein}}, \ and\ \bibinfo {author}
  {\bibfnamefont {H.}~\bibnamefont {Stoecker}},\ }\href {\doibase
  10.1103/PhysRevC.102.035202} {\bibfield  {journal} {\bibinfo  {journal}
  {Phys. Rev. C}\ }\textbf {\bibinfo {volume} {102}},\ \bibinfo {pages}
  {035202} (\bibinfo {year} {2020})}\BibitemShut {NoStop}%
\bibitem [{\citenamefont {Fernández-Ramírez}\ \emph
  {et~al.}(2018)\citenamefont {Fernández-Ramírez}, \citenamefont {Lo},\ and\
  \citenamefont {Petreczky}}]{Fernandez-Ramirez:2018vzu}%
  \BibitemOpen
  \bibfield  {author} {\bibinfo {author} {\bibfnamefont {C.}~\bibnamefont
  {Fernández-Ramírez}}, \bibinfo {author} {\bibfnamefont {P.~M.}\
  \bibnamefont {Lo}}, \ and\ \bibinfo {author} {\bibfnamefont {P.}~\bibnamefont
  {Petreczky}},\ }\href {\doibase 10.1103/PhysRevC.98.044910} {\bibfield
  {journal} {\bibinfo  {journal} {Phys. Rev. C}\ }\textbf {\bibinfo {volume}
  {98}},\ \bibinfo {pages} {044910} (\bibinfo {year} {2018})},\ \Eprint
  {http://arxiv.org/abs/1806.02177} {arXiv:1806.02177 [hep-ph]} \BibitemShut
  {NoStop}%
\bibitem [{\citenamefont {Dash}\ \emph {et~al.}(2019)\citenamefont {Dash},
  \citenamefont {Samanta},\ and\ \citenamefont {Mohanty}}]{Dash:2018mep}%
  \BibitemOpen
  \bibfield  {author} {\bibinfo {author} {\bibfnamefont {A.}~\bibnamefont
  {Dash}}, \bibinfo {author} {\bibfnamefont {S.}~\bibnamefont {Samanta}}, \
  and\ \bibinfo {author} {\bibfnamefont {B.}~\bibnamefont {Mohanty}},\ }\href
  {\doibase 10.1103/PhysRevC.99.044919} {\bibfield  {journal} {\bibinfo
  {journal} {Phys. Rev. C}\ }\textbf {\bibinfo {volume} {99}},\ \bibinfo
  {pages} {044919} (\bibinfo {year} {2019})},\ \Eprint
  {http://arxiv.org/abs/1806.02117} {arXiv:1806.02117 [hep-ph]} \BibitemShut
  {NoStop}%
\bibitem [{\citenamefont {Anchishkin}\ \emph {et~al.}(2019)\citenamefont
  {Anchishkin}, \citenamefont {Mishustin},\ and\ \citenamefont
  {Stoecker}}]{Anchishkin_2019}%
  \BibitemOpen
  \bibfield  {author} {\bibinfo {author} {\bibfnamefont {D.}~\bibnamefont
  {Anchishkin}}, \bibinfo {author} {\bibfnamefont {I.}~\bibnamefont
  {Mishustin}}, \ and\ \bibinfo {author} {\bibfnamefont {H.}~\bibnamefont
  {Stoecker}},\ }\href {\doibase 10.1088/1361-6471/aafea8} {\bibfield
  {journal} {\bibinfo  {journal} {Journal of Physics G: Nuclear and Particle
  Physics}\ }\textbf {\bibinfo {volume} {46}},\ \bibinfo {pages} {035002}
  (\bibinfo {year} {2019})}\BibitemShut {NoStop}%
\bibitem [{\citenamefont {Walecka}(1974)}]{Walecka:1974qa}%
  \BibitemOpen
  \bibfield  {author} {\bibinfo {author} {\bibfnamefont {J.~D.}\ \bibnamefont
  {Walecka}},\ }\href {\doibase 10.1016/0003-4916(74)90208-5} {\bibfield
  {journal} {\bibinfo  {journal} {Annals Phys.}\ }\textbf {\bibinfo {volume}
  {83}},\ \bibinfo {pages} {491} (\bibinfo {year} {1974})}\BibitemShut
  {NoStop}%
\bibitem [{\citenamefont {Landau}\ and\ \citenamefont {Lifshitz}(1975)}]{LL}%
  \BibitemOpen
  \bibfield  {author} {\bibinfo {author} {\bibfnamefont {L.~D.}\ \bibnamefont
  {Landau}}\ and\ \bibinfo {author} {\bibfnamefont {E.~M.}\ \bibnamefont
  {Lifshitz}},\ }\href@noop {} {\emph {\bibinfo {title} {Statistical
  Physics}}}\ (\bibinfo  {publisher} {Pergamon, Oxford},\ \bibinfo {year}
  {1975})\BibitemShut {NoStop}%
\bibitem [{\citenamefont {Poberezhnyuk}\ \emph {et~al.}(2017)\citenamefont
  {Poberezhnyuk}, \citenamefont {Vovchenko}, \citenamefont {Anchishkin},\ and\
  \citenamefont {Gorenstein}}]{Poberezhnyuk:2017yhx}%
  \BibitemOpen
  \bibfield  {author} {\bibinfo {author} {\bibfnamefont {R.}~\bibnamefont
  {Poberezhnyuk}}, \bibinfo {author} {\bibfnamefont {V.}~\bibnamefont
  {Vovchenko}}, \bibinfo {author} {\bibfnamefont {D.}~\bibnamefont
  {Anchishkin}}, \ and\ \bibinfo {author} {\bibfnamefont {M.}~\bibnamefont
  {Gorenstein}},\ }\href {\doibase 10.1142/S0218301317500616} {\bibfield
  {journal} {\bibinfo  {journal} {Int.\ J.\ Mod.\ Phys.\ E}\ }\textbf {\bibinfo
  {volume} {26}},\ \bibinfo {pages} {1750061} (\bibinfo {year} {2017})},\
  \Eprint {http://arxiv.org/abs/1708.05605} {arXiv:1708.05605 [nucl-th]}
  \BibitemShut {NoStop}%
\bibitem [{\citenamefont {Stashko}\ \emph
  {et~al.}(2021{\natexlab{b}})\citenamefont {Stashko}, \citenamefont {Savchuk},
  \citenamefont {Poberezhnyuk}, \citenamefont {Vovchenko},\ and\ \citenamefont
  {Gorenstein}}]{PhysRevC.103.065201}%
  \BibitemOpen
  \bibfield  {author} {\bibinfo {author} {\bibfnamefont {O.~S.}\ \bibnamefont
  {Stashko}}, \bibinfo {author} {\bibfnamefont {O.~V.}\ \bibnamefont
  {Savchuk}}, \bibinfo {author} {\bibfnamefont {R.~V.}\ \bibnamefont
  {Poberezhnyuk}}, \bibinfo {author} {\bibfnamefont {V.}~\bibnamefont
  {Vovchenko}}, \ and\ \bibinfo {author} {\bibfnamefont {M.~I.}\ \bibnamefont
  {Gorenstein}},\ }\href {\doibase 10.1103/PhysRevC.103.065201} {\bibfield
  {journal} {\bibinfo  {journal} {Phys. Rev. C}\ }\textbf {\bibinfo {volume}
  {103}},\ \bibinfo {pages} {065201} (\bibinfo {year}
  {2021}{\natexlab{b}})}\BibitemShut {NoStop}%
\bibitem [{\citenamefont {Greiner}\ \emph {et~al.}(2012)\citenamefont
  {Greiner}, \citenamefont {Neise},\ and\ \citenamefont {St{\"o}cker}}]{GNS}%
  \BibitemOpen
  \bibfield  {author} {\bibinfo {author} {\bibfnamefont {W.}~\bibnamefont
  {Greiner}}, \bibinfo {author} {\bibfnamefont {L.}~\bibnamefont {Neise}}, \
  and\ \bibinfo {author} {\bibfnamefont {H.}~\bibnamefont {St{\"o}cker}},\
  }\href@noop {} {\emph {\bibinfo {title} {Thermodynamics and statistical
  mechanics}}}\ (\bibinfo  {publisher} {Springer Science \& Business Media},\
  \bibinfo {year} {2012})\BibitemShut {NoStop}%
\bibitem [{\citenamefont {Anchishkin}\ and\ \citenamefont
  {Vovchenko}(2015)}]{Anchishkin_2015}%
  \BibitemOpen
  \bibfield  {author} {\bibinfo {author} {\bibfnamefont {D.}~\bibnamefont
  {Anchishkin}}\ and\ \bibinfo {author} {\bibfnamefont {V.}~\bibnamefont
  {Vovchenko}},\ }\href {\doibase 10.1088/0954-3899/42/10/105102} {\bibfield
  {journal} {\bibinfo  {journal} {Journal of Physics G: Nuclear and Particle
  Physics}\ }\textbf {\bibinfo {volume} {42}},\ \bibinfo {pages} {105102}
  (\bibinfo {year} {2015})}\BibitemShut {NoStop}%
\bibitem [{\citenamefont {Brandt}\ \emph
  {et~al.}(2018{\natexlab{d}})\citenamefont {Brandt}, \citenamefont
  {Endrődi},\ and\ \citenamefont {Schmalzbauer}}]{BBrant_2018}%
  \BibitemOpen
  \bibfield  {author} {\bibinfo {author} {\bibfnamefont {B.~B.}\ \bibnamefont
  {Brandt}}, \bibinfo {author} {\bibfnamefont {G.}~\bibnamefont {Endrődi}}, \
  and\ \bibinfo {author} {\bibfnamefont {S.}~\bibnamefont {Schmalzbauer}},\
  }\href {\doibase 10.1051/epjconf/201817507020} {\bibfield  {journal}
  {\bibinfo  {journal} {EPJ Web of Conferences}\ }\textbf {\bibinfo {volume}
  {175}},\ \bibinfo {pages} {07020} (\bibinfo {year}
  {2018}{\natexlab{d}})}\BibitemShut {NoStop}%
\bibitem [{\citenamefont {Brandt}\ and\ \citenamefont {Endrődi}(2019)}]{2019}%
  \BibitemOpen
  \bibfield  {author} {\bibinfo {author} {\bibfnamefont {B.}~\bibnamefont
  {Brandt}}\ and\ \bibinfo {author} {\bibfnamefont {G.}~\bibnamefont
  {Endrődi}},\ }\href {\doibase 10.1103/physrevd.99.014518} {\bibfield
  {journal} {\bibinfo  {journal} {Physical Review D}\ }\textbf {\bibinfo
  {volume} {99}} (\bibinfo {year} {2019}),\
  10.1103/physrevd.99.014518}\BibitemShut {NoStop}%
\bibitem [{\citenamefont {Karsch}\ and\ \citenamefont
  {Redlich}(2011)}]{Karsch_2011}%
  \BibitemOpen
  \bibfield  {author} {\bibinfo {author} {\bibfnamefont {F.}~\bibnamefont
  {Karsch}}\ and\ \bibinfo {author} {\bibfnamefont {K.}~\bibnamefont
  {Redlich}},\ }\href {\doibase 10.1016/j.physletb.2010.10.046} {\bibfield
  {journal} {\bibinfo  {journal} {Physics Letters B}\ }\textbf {\bibinfo
  {volume} {695}},\ \bibinfo {pages} {136–142} (\bibinfo {year}
  {2011})}\BibitemShut {NoStop}%
\end{thebibliography}%

\end{document}